\documentclass[reprint,superscriptaddress,amsmath,amssymb,aps,prb,longbibliography]{revtex4-2}
\usepackage{graphics,graphicx,epsfig,color,latexsym,float,amsmath,bm}
\usepackage[colorlinks,citecolor=blue,linkcolor=blue,urlcolor=blue]{hyperref}
\usepackage{xurl}

\usepackage[version=4]{mhchem}
\usepackage{wrapfig}
\usepackage{ulem}
\usepackage{lipsum}
\usepackage{placeins}
\usepackage{braket}
\usepackage{appendix}
\usepackage{booktabs}

\begin{document}

\title{Engineering non-ergodic properties in two dimensional quantum many-body systems}

\author{Nyayabanta Swain}
\affiliation{Centre for Quantum Technologies, National University of Singapore, Singapore 117543}
\affiliation{Department of Materials Science and Engineering, National University of Singapore, 9 Engineering Drive 1, Singapore 117575}

\author{\mbox{Gabriel Lemari\'e}}
\affiliation{INPHYNI, Universit\'e C\^ote d'Azur, CNRS, Nice, France}

\author{Shaffique Adam}
\affiliation{Department of Materials Science and Engineering, National University of Singapore, 9 Engineering Drive 1, Singapore 117575}
\affiliation{Department of Physics, Washington University in St. Louis, St. Louis, Missouri 63130, United States}

\begin{abstract}
Non-ergodic quantum many-body dynamics offers a route to persistent quantum coherence far from equilibrium, beyond the conventional expectations of thermalizing statistical mechanics. Most known examples have been identified through intuition, analogy, or numerical search, rather than by systematic Hamiltonian design.  
This leaves few general methods for engineering robust non-ergodic dynamics in disorder-free interacting quantum many-body systems, especially in higher dimensions where enhanced connectivity generally disfavors ergodicity breaking.  Here, we use an eigenstate-to-Hamiltonian construction approach to systematically engineer two-dimensional quantum spin Hamiltonians with tunable non-ergodic properties. Starting from a common parent model and input eigenstate, we construct two target Hamiltonians on the square lattice that 
differ in their interaction geometry: axial versus diagonal. 
We show that the resulting interaction geometry and coupling pattern control the Hilbert-space structure and dynamics.  The axial model forms a connected Hilbert-space network and exhibits an anisotropy-driven crossover from ergodic to non-ergodic behavior through the suppression of resonances. In contrast, the diagonal model exhibits Hilbert-space fragmentation and, within its largest irreducible sector, signatures of quantum many-body scar dynamics including towers of highly localized, low-entanglement eigenstates, and long-lived coherent revivals. These results establish a direct link between Hamiltonian design, Hilbert-space network structure, and emergent non-ergodic dynamics.
\end{abstract}
\maketitle

\section{Introduction}
\label{sec_intro}
Understanding how isolated quantum many-body systems thermalize remains a central problem in nonequilibrium physics~\cite{Thermalization_Rigol_2008}. 
In closed systems, unitary evolution precludes coupling to an external bath, implying that thermal behavior must emerge from intrinsic many-body dynamics~\cite{ETH_review_2016}. 
Nevertheless, a wide class of interacting quantum systems exhibit robust thermalization, where expectation values of local observables relax toward values described by statistical mechanics, even when the system evolves from a far from equilibrium initial state. This phenomenon is elegantly described by the eigenstate thermalization hypothesis (ETH)~\cite{ETH_1991,ETH_1994,ETH_review_2016,Deutsch_2018}. ETH provides a conceptual bridge between microscopic quantum dynamics and macroscopic statistical mechanics and has been extensively verified in quantum chaotic many-body systems~\cite{ETH_review_2016,Deutsch_2018}.

While ETH successfully explains thermalization in many interacting systems, it is now well established that it is not always valid~\cite{DAlessio2016,Abanin2019}. There are known cases where quantum systems fail to thermalize and exhibit ergodicity breaking, where the eigenstates retain nonthermal structure and long-time dynamics remain sensitive to initial conditions. Several distinct mechanisms are known to lead to violations of ETH~\cite{DAlessio2016}. 
Historically, the most familiar mechanism of ergodicity breaking is integrability, where an extensive set of conserved quantities constrains the dynamics~\cite{Korepin1993,Faddeev1996,integrable1,integrable2,Faddeev2013,integrable3,Rev_Quantum_chaos_2026}. 
A different route to non-ergodicity arises in disordered systems, where Anderson localization and its interacting counterpart, many-body localization (MBL), prevent thermalization entirely~\cite{Nandkishore2014,Alet2018,Abanin2019,GOPALAKRISHNAN20201,sierant2024manybody}.
More recently, it has been recognized that even without disorder, interacting systems can evade thermalization. In quantum many-body scars (QMBS), kinetic constraints produce a small subset of atypical eigenstates embedded in an otherwise thermal spectrum, resulting in weak ergodicity breaking~\cite{rydberg_atom_2017,scar_states_Turner2018,PhysRevLett.122.173401,scar_states_Serbyn2021,rev_scars_2023,Pizzi2025QuantumScars,ScarFinder_Papic2025}.  A second mechanism is Hilbert-space fragmentation (HSF), where additional symmetries and dynamical constraints split the Hilbert space into disconnected sectors~\cite{HSF_Sala_PRX2020,HSF_local_PRB_2020,PhysRevX.12.011050,scar_rev_Moudgalya_2022}.  These phenomena have been extensively explored in one-dimensional systems, where both analytical and numerical methods provide detailed insight~\cite{integrable1,Pal2010,Znidaric2008MBL,Luitz2015,KT_MBL_Gabriel2020,scar_states_Turner2018,Surace2021Scars,Brenes2020ETH,PhysRevX.10.011055,You2022KitaevScars}.
However, in higher dimensions our understanding remains comparatively limited~\cite{Chandran2016,Wahl2017TensorMBL,Potirniche2019,Gopalakrishnan2019Instability,Doggen2020,Alet_2D-loc_dimer_2020,MBL2DTopological2024}. This raises a natural question $-$ \textit{can such non-ergodic phenomena be realized in two dimensional quantum systems, and can these be identified systematically?} 
To address this question, we employ the eigenstate-to-Hamiltonian construction (EHC)~\cite{Chertkov2018,Qi2019,Dupont2019a,Dupont2019b}, which provides a route to engineer two dimensional Hamiltonians with at least one non-ergodic high-energy excited eigenstate. 

The EHC approach provides a systematic way to build many-body models by reversing the conventional approach of solving for eigenstates of a given Hamiltonian. EHC begins with a chosen many-body state and asks what Hamiltonians can host it as an eigenstate. Given an input state \(\ket{\Psi}\) and a set of local operators \(\{\mathcal O_i\} \), one constructs a covariance matrix whose eigenvectors determine the target Hamiltonians for which the state has minimal energy variance~\cite{Chertkov2018}. If the variance vanishes, the state is an exact eigenstate of the target Hamiltonian~\cite{Chertkov2018,Qi2019}. If the variance is non-zero, but small, and vanishes in the thermodynamic limit, then the chosen eigenstate is an approximate eigenstate of the target Hamiltonian~\cite{Dupont2019a,Swain_EHC_QMC_PRB_2025}.
EHC is therefore useful in this setting because it can generate target Hamiltonians containing atypical eigenstates, which may seed broader non-ergodic dynamics.

We begin from a ground state that is itself non-ergodic.
Through the EHC mapping, this state is promoted to a highly excited eigenstate of a new target Hamiltonian. This immediately implies the existence of at least one non-ergodic eigenstate in the spectrum. As we demonstrate below, the physics is much richer --- the target Hamiltonian hosts not just a single such state, but a large number of atypical eigenstates exhibiting pronounced non-ergodic behavior. In this way, the EHC framework naturally generates models containing atypical low-entanglement eigenstates. More importantly, the computational complexity of the EHC method depends on the number of operators included in the construction rather than on the exponentially large Hilbert-space dimension, making it a practical tool when suitable operator bases are chosen.

\begin{figure*}[t]
\centerline{
\includegraphics[width=0.98\linewidth]{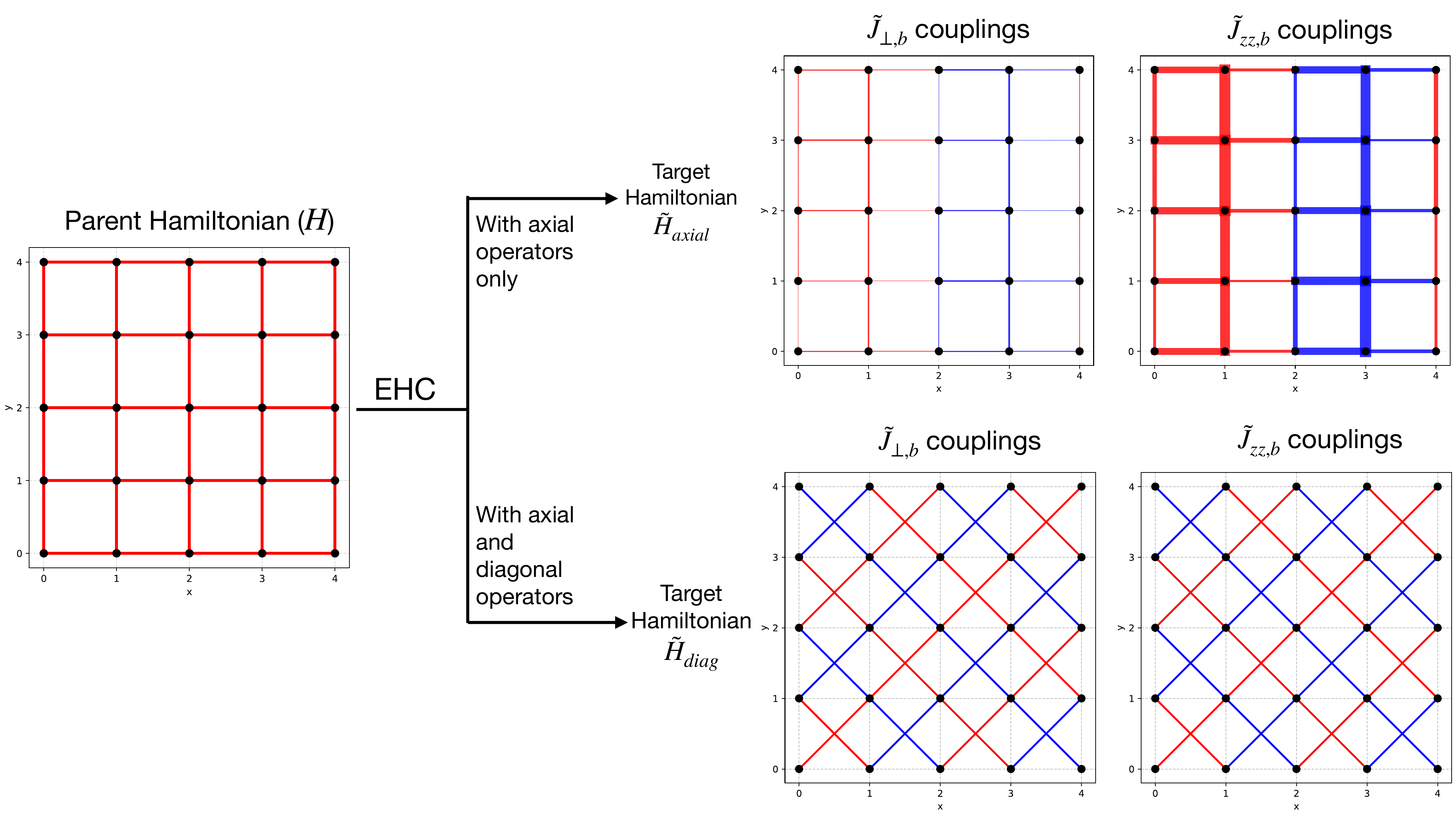}
}
\caption{
\textbf{Visual representation of the couplings of the parent and target Hamiltonians:}
This figure illustrates the construction of distinct two-dimensional target Hamiltonians obtained from a common parent Hamiltonian using the eigenstate-to-Hamiltonian construction approach.
The parent Hamiltonian, $H$ is the spin-1/2 XXZ model with antiferromagnetic spin-flip and spin preserving couplings of unit magnitude (shown as red bonds) between nearest-neighbor (NN) spins and no couplings along the  next-nearest-neighbor (NNN) bonds. 
Applying EHC with only axial (NN) operators yields the target Hamiltonian \(\tilde{H}_{\mathrm{axial}}\) where the couplings \(\tilde{J}_{\perp,b}\) and \(\tilde{J}_{zz,b}\), exhibit a characteristic \((\pi/2,0)\) ordering in momentum space. In real space (upper panel), the couplings are represented as sticks of varying thickness proportional to their magnitude, and with red and blue colors indicating antiferromagnetic and ferromagnetic bonds, respectively. It is worth noting that \( |\tilde{J}_{zz,b}| =6|\tilde{J}_{\perp,b}|\) on each bond for $\Delta = 4$.
Including both axial (NN) and diagonal (NNN) operators in the EHC basis produces the second target Hamiltonian \(\tilde{H}_{\mathrm{diag}}\) where NN couplings vanish and NNN couplings remain finite (see lower panel). The \(\tilde{J}_{\perp,b}\) and \(\tilde{J}_{zz,b}\) couplings on diagonal bonds display a \((\pi,\pi)\) ordering in momentum space. In real space, the couplings change from antiferromagnetic to ferromagnetic on alternating plaquettes with \(|\tilde{J}_{zz,b}| \approx 1.18 |\tilde{J}_{\perp,b}|\) on each bond for $\Delta = 4$.} 
\label{fig:1}
\end{figure*} 

\section{State-of-the-art of known mechanisms of non-ergodicity}
\label{sec_state_of_art}

We first outline the principal mechanisms through which ergodicity can be violated in interacting quantum systems. These include mechanisms such as integrability, many-body localization, and Hilbert-space fragmentation. There also exist weaker forms of ergodicity breaking associated with quantum many-body scars. Each mechanism leaves distinct signatures in spectral properties, eigenstate structure, and real-time dynamics.

Integrable systems represent the simplest and most well-understood examples of non-ergodic dynamics~\cite{Korepin1993,Faddeev1996,Thermalization_Rigol_2008}. Such systems possess an extensive number of mutually commuting conserved quantities, whose number grows with system size~\cite{Faddeev2013,integrable3}. These conservation laws severely restrict the dynamics and prevent conventional thermalization. The spectral properties of integrable systems reflect this constrained dynamics; their energy levels follow Poisson statistics, indicating the absence of level repulsion between eigenstates~\cite{Rev_Quantum_chaos_2026}. 
In contrast, non-integrable systems typically exhibit quantum chaos, characterized by strong hybridization among eigenstates and spectral statistics described by random matrix theory~\cite{Rev_Quantum_chaos_2026}. In this regime, highly excited eigenstates satisfy ETH and isolated systems relax to thermal equilibrium under unitary evolution.

The transition between these regimes occurs when integrability-breaking perturbations are introduced. Classical Hamiltonian systems provide valuable intuition through the Kolmogorov–Arnold–Moser theorem, which describes the gradual destruction of regular trajectories as perturbations increase~\cite{arnold1989mathematical, kolmogorov1954preservation,arnold1963proof,moser1962invariant}. At weak perturbations, large regions of regular motion coexist with small chaotic regions in phase space. As perturbations grow, chaotic regions expand until global chaos emerges~\cite{PhysRevResearch.5.043019, chirikov1979universal, lichtenberg1992regular}. Upon quantization, this mixed phase-space structure leads to the coexistence of eigenstates with markedly different properties $-$ states associated with regular trajectories remain weakly hybridized, while states associated with chaotic regions can be delocalized on the chaotic sea \cite{bohigas1993manifestations, Ott_2002} or display dynamical localization \cite{casati1979stochastic, santhanam2022quantum}. Consequently, even weakly nonintegrable systems can exhibit remnants of non-ergodic behavior over large portions of the spectrum before eventually crossing over to fully chaotic dynamics~\cite{PhysRevB.104.184302}.

A qualitatively different mechanism for ergodicity breaking arises from disorder-induced localization. In non-interacting systems, strong disorder leads to Anderson localization, signaling the localization of single-particle wave functions through destructive quantum interference, and suppressing transport~\cite{Anderson_1958,anderson_loc_RMP_2008,abrahams2010fifty}.
In interacting systems, localization can persist in the form of MBL~\cite{Nandkishore2014,Alet2018,Abanin2019,GOPALAKRISHNAN20201,sierant2024manybody}. In the MBL phase, strong quenched disorder prevents the system from acting as its own thermal bath, resulting in a breakdown of ETH across the entire spectrum. Eigenstates remain non-ergodic in the configuration space and obey an area-law entanglement scaling even at finite energy density. As a consequence, MBL systems retain memory of their initial conditions for arbitrarily long times and exhibit highly nonergodic dynamics. MBL therefore represents a striking example of ergodicity breaking in interacting systems, driven not by conserved quantities but by disorder and quantum interference.

Another intriguing form of ETH violation is known as Hilbert-space fragmentation~\cite{HSF_Sala_PRX2020,HSF_local_PRB_2020}. In fragmented systems, the many-body Hilbert space decomposes into exponentially many dynamically disconnected sectors such that the Hamiltonian has no matrix elements connecting states belonging to different sectors. 
Unlike integrability, HSF does not arise from conventional conserved quantities. Instead, it typically emerges from local dynamical constraints or emergent symmetries that restrict the allowed transitions between configurations.
Operationally, fragmentation can be revealed through Krylov subspace constructions, where repeated action of the Hamiltonian on a product state generates a subspace that remains exponentially smaller than the full Hilbert space~\cite{PhysRevX.12.011050,scar_rev_Moudgalya_2022}. Two types of fragmentation are commonly distinguished. In strong fragmentation, even the largest connected sector occupies a vanishing fraction of the Hilbert space in the thermodynamic limit~\cite{PhysRevLett.124.207602}. In weak fragmentation, the largest sector remains thermodynamically large and typically thermalizes, while smaller sectors exhibit atypical dynamics~\cite{scar_rev_Moudgalya_2022,Loh_expt_rydberg_PRX_2025}.

Recent works have uncovered another mechanism for ergodicity breaking in interacting quantum many-body systems without disorder known as QMBS~\cite{rydberg_atom_2017,scar_states_Turner2018,PhysRevLett.122.173401,scar_states_Serbyn2021,rev_scars_2023,Pizzi2025QuantumScars}. In these  systems, the majority of eigenstates satisfy the ETH, yet a small set of atypical eigenstates embedded within the thermal spectrum remain nonthermal. These special scar states can dominate the dynamics when the system is initialized in particular product states, leading to long-lived coherent oscillations and revivals instead of rapid thermalization. Scar states often form nearly equally spaced towers of eigenstates and typically arise in systems with kinetic constraints or special symmetry structures~\cite{scar_states_Turner2018,scar_states_Turner2018_2}. Unlike integrability or localization, QMBS represent weak ergodicity breaking, where only a small fraction of the spectrum violates ETH while the remainder behaves thermally.
The first experimental evidence for QMBS is observed in Rydberg atom arrays, where strong blockade interactions prevent simultaneous excitation of neighboring atoms~\cite{rydberg_atom_2017}. Certain initial states in these systems exhibit long-lived periodic revivals rather than the rapid thermalization expected for a generic interacting model.
These dynamics can be captured by the PXP model, a constrained spin-1/2 Hamiltonian that reproduces the Rydberg blockade mechanism~\cite{scar_states_Turner2018,PhysRevLett.122.173401,scar_states_Serbyn2021}. Within its spectrum, a small set of special eigenstates forms a nearly equally spaced tower and exhibits large overlap with particular product states, enabling coherent many-body oscillations.
Subsequent studies have identified scars in a wide variety of models, including spin, bosonic, and fermionic systems~\cite{scar_2018_Moudgalya,Schecter_scar_XY_2019,scars_2019_Bull,scars_2020_Iadecola,Khemani_scar_XY_2020,WenWei_scar_XY_2020,scar_disorder_spin-chain_2020,PhysRevB.102.224303,Swain_EHC_QMC_PRB_2025}. Experiments have also realized scar dynamics in driven Rydberg arrays and superconducting quantum processors~\cite{scar_expt_driven_rydberg_2021,Zhang_scar_expt_qcomp_2023,Gustafson_scar_qcomp_2023}.
Although most theoretical work has focused on one-dimensional systems, recent advances in programmable quantum simulators have enabled studies of scar physics in two dimensions, particularly in Rydberg-atom arrays with tunable geometries~\cite{scar_expt_driven_rydberg_2021}. These developments raise new questions about the nature and stability of scar dynamics in higher dimensions.

\section{Summary of main results}

Using EHC, we engineer two classes of two-dimensional target Hamiltonians, \( \tilde{H}_{\mathrm{axial}} \) and \( \tilde{H}_{\mathrm{diag}} \),
starting from the same input eigenstate but using different operator bases, which generate distinct interaction geometries and coupling patterns. As seen in Fig.~\ref{fig:1}, the axial model contains nearest-neighbor (NN) couplings along the horizontal and vertical directions of the square lattice, whereas the diagonal model contains only next-nearest-neighbor (NNN) couplings along the lattice diagonals.  Although the two models share the same EHC origin, their generated coupling patterns differ sharply in both bond geometry and spatial order. We find that the resulting Hamiltonians exhibit strikingly different Hilbert-space structure and non-ergodic behavior, demonstrating that interaction geometry can dictate the emergence of ergodicity breaking in our EHC-engineered models.

For \( \tilde{H}_{\mathrm{axial}}\) , we find a crossover from ergodic to strongly non-ergodic behavior driven by the suppression of resonances with increasing anisotropy \(\Delta \). Resonance analysis reveals that the Hilbert-space network evolves from a densely resonant structure at small \(\Delta \) to a more sparser regime at large \( \Delta \). Correspondingly, the eigenspectrum develops strong inhomogeneity, with highly localized eigenstates coexisting alongside more delocalized states even in the middle of the spectrum. Spectral diagnostics based on inverse participation ratio (IPR) and entanglement entropy (EE) show the gradual reduction of ergodic behavior and the emergence of strongly non-ergodic eigenstates. Dynamically, Gaussian wave packets rapidly dephase without exhibiting any coherent revivals, while Fock-space evolution reveals qualitatively different spreading behavior depending on the initial basis state, reflecting the highly inhomogeneous structure of the Hilbert space graph.

In contrast, \(\tilde{H}_{\mathrm{diag}} \) realizes a different mechanism of weak ergodicity breaking. Here the Hilbert space fragments into several disconnected sectors with strongly heterogeneous resonance structures. Some fragmented blocks exhibit dense resonances and more ergodic behavior, while others remain weakly resonant and strongly non-ergodic. This gives rise to pronounced tower-like structures in both IPR and EE, consisting of atypical low-entangled, high-IPR eigenstates coexisting with a broader ergodic background. These structured eigenstate towers give rise to coherent revivals in Gaussian wave-packet dynamics and recurrent Fock-space evolution. This suggests the presence of quantum many-body scar-like dynamics associated with the fragmented Hilbert-space structure and heterogeneous resonance patterns. Taken together, our results show how interaction geometry shapes Hilbert-space connectivity and resonance structure, giving rise to distinct pathways to ergodicity breaking in two-dimensional disorder-free quantum systems.

\section{Model and Method}
\label{sec_model_method}

We begin with the $S=1/2$ antiferromagnetic XXZ model in two dimensions with periodic boundary conditions,
\begin{align}
H= \sum_{\langle i,j \rangle} [ (J/2) (S_i^{+}  S_{j}^{-} + S_i^{-}  S_{j}^{+}) + \Delta J S_i^z S_{j}^z ]
\label{eq1}
\end{align}
where $S^{+(-)}_i$ is the raising (lowering) operator for a spin at site $i$, $S^z_i$ is the projection of that spin on the $z$-axis. \(J\) is the strength of antiferromagnetic coupling for nearest-neighbor sites along the x and y directions, while \(\Delta\) is the anisotropy parameter of the 2D XXZ model.

We compute the ground state $|\Psi_{\mathrm{GS}}\rangle$ of the parent Hamiltonian \( H \). Next, we conduct a search for a target Hamiltonian~\cite{Dupont2019a,Dupont2019b}, with a form similar to \( H \) as in Eq. (\ref{eq1}), and with finite interactions between the spins along the nearest-neighbor spins along the x and y directions, but no interactions for next-nearest-neighbor sites along the diagonal directions. We denote this target Hamiltonian as \( \tilde{H}_{\mathrm{axial}} \).

To construct \(\tilde{H}_{\mathrm{axial}}\) within the EHC framework, we consider nearest-neighbor (NN) bonds $b=\langle i,j\rangle$ on a square lattice of size $L\times L$. For each bond, we define two local operators $-$ a spin-exchange interaction term and an Ising interaction term. Collectively, this yields a set of $2N_b$ basis operators, where $N_b=2L^2$ is the total number of NN bonds on the lattice.
Specifically, for each bond $b=\langle i,j\rangle$, we define $\mathcal{O}_{b}^{(1)}=\tfrac{1}{2}(S_i^{+}S_j^{-}+S_i^{-}S_j^{+})$, and $\mathcal{O}_{b}^{(2)}=S_i^{z}S_j^{z}$.
Using this operator basis $\{\mathcal{O}_\alpha\}$ with $\alpha=1,\dots,2N_b$, we construct the covariance matrix
$\mathcal{C}_{\alpha\beta}
= \langle \Psi_{\mathrm{GS}} | \mathcal{O}_\alpha \mathcal{O}_\beta | \Psi_{\mathrm{GS}} \rangle
- \langle \Psi_{\mathrm{GS}} | \mathcal{O}_\alpha | \Psi_{\mathrm{GS}} \rangle
\langle \Psi_{\mathrm{GS}} | \mathcal{O}_\beta | \Psi_{\mathrm{GS}} \rangle$,
where $|\Psi_{\mathrm{GS}}\rangle$ is the reference ground state.
The resulting covariance matrix has dimension $2N_b \times 2N_b$.

We diagonalize the covariance matrix and find that its lowest eigenvalue is exactly zero. The corresponding eigenvector defines the coupling parameters of the parent Hamiltonian. The next four eigenvalues are degenerate and vanishingly small, indicating the presence of a low-dimensional subspace of target Hamiltonians with quasi-zero modes. We choose a superposition (equal weight, but arbitrary linear combination) of these four eigenvectors to perform the EHC procedure. Since the eigenvalues of the covariance matrix correspond to energy variances, the eigenvectors associated with vanishing eigenvalues encode the parameters of the target Hamiltonians. This leads to the systematic generation of a family of target Hamiltonians given by
\begin{equation}
\tilde{H}_{\mathrm{axial}} = \sum_{\langle i,j \rangle =NN} [ \tfrac{\tilde{J}_{\perp,b} }{2} (S_i^{+}  S_{j}^{-} + S_i^{-}  S_{j}^{+}) + \Delta \tilde{J}_{zz,b} S_i^z S_{j}^z ] 
\label{eq2}
\end{equation}

Upon investigation, we find that the couplings, \(\tilde{J}_{\perp,b}\) and \(\tilde{J}_{zz,b}\), exhibit a characteristic \((\pi/2,0)\) or \((0, \pi/2)\)order in the momentum space. The different supperpositions give either one of these ordering. 
In fig.~\ref{fig:1}, we show the \((\pi/2,0)\) ordering, with the couplings in real space denoted as sticks whose thickness is proportional to their magnitude, and with red and blue colors indicating the antiferromagnetic and ferromagnetic interactions, respectively. As seen in the upper panel, the couplings along x-direction change from antiferromagnetic (red) to ferromagnetic (blue) after two lattice spacings, whereas the couplings along the y-direction remain the same. We also find that \( |\tilde{J}_{zz,b}| =6|\tilde{J}_{\perp,b}|\) on each bond for $\Delta = 4$. In Appendix \ref{app_B}, we present more details of the $\Delta$ dependence of \( |\tilde{J}_{zz,b}|/|\tilde{J}_{\perp,b}|\).

Next, extend the EHC operator basis to include next-nearest-neighbor interactions along the diagonal bonds of the square lattice. In addition to the $2N_b$ operators defined on nearest-neighbor bonds, we introduce an additional $2N_b$ operators acting on diagonal bonds, resulting in a total of $4N_b$ basis operators.
Labeling NN bonds by $b=\langle i,j \rangle$ and NNN bonds by $b=\langle\langle i,j \rangle\rangle$, the operator set is defined as
$\mathcal{O}_b^{(\perp)} = \tfrac{1}{2}\left(S_i^{+} S_j^{-} + S_i^{-} S_j^{+}\right)$, and
$\mathcal{O}_b^{(z)} = S_i^{z} S_j^{z}$,
where $b$ runs over all $2N_b$ bonds (NN and NNN). 
The full operator basis is then $\{\mathcal{O}_\alpha\}_{\alpha=1}^{4N_b}$, combining spin-exchange and Ising interaction terms on each bond.
Using this expanded basis, we construct the covariance matrix
$\mathcal{C}_{\alpha\beta}=
\langle \Psi_{\mathrm{GS}} | \mathcal{O}_\alpha \mathcal{O}_\beta | \Psi_{\mathrm{GS}} \rangle
-
\langle \Psi_{\mathrm{GS}} | \mathcal{O}_\alpha | \Psi_{\mathrm{GS}} \rangle
\langle \Psi_{\mathrm{GS}} | \mathcal{O}_\beta | \Psi_{\mathrm{GS}} \rangle$,
which now has dimension $4N_b \times 4N_b$.

Diagonalization of this enlarged covariance matrix reveals non-degenerate eigenvectors corresponding to vanishing eigenvalues. These identify possible target Hamiltonians within the expanded operator space. Of particular interest are the solutions where the dominant couplings occur along the diagonal NNN bonds, while the NN interactions along the x and y directions are effectively suppressed. These eigenvectors define a new class of target Hamiltonians  given by
\begin{equation}
\tilde{H}_{\mathrm{diag}} = \sum_{\langle\langle i,j \rangle\rangle =NNN} [ \tfrac{\tilde{J}_{\perp,b} }{2} (S_i^{+}  S_{j}^{-} + S_i^{-}  S_{j}^{+}) + \Delta \tilde{J}_{zz,b} S_i^z S_{j}^z ] 
\label{eq3}
\end{equation}

We find that the couplings, \(\tilde{J}_{\perp,b}\) and \(\tilde{J}_{zz,b}\), of \(\tilde{H}_{\mathrm{diag}}\), vanish on NN bonds but remain finite on NNN bonds. These couplings display a characteristic order \((\pi,\pi)\) in the momentum space. In fig.~\ref{fig:1}, lower panel, we show the couplings in real space. The couplings change from antiferromagnetic to ferromagnetic on alternating plaquettes of the square lattice with \(|\tilde{J}_{zz,b}| \approx 1.18 |\tilde{J}_{\perp,b}|\) on each bond for $\Delta = 4$. Appendix \ref{app_B} has further details of the $\Delta$ dependence of \( |\tilde{J}_{zz,b}|/|\tilde{J}_{\perp,b}|\).

\section{Results and Discussion}
\label{sec_results}

\begin{figure}[t]
\centerline{
\includegraphics[width=0.95\linewidth]{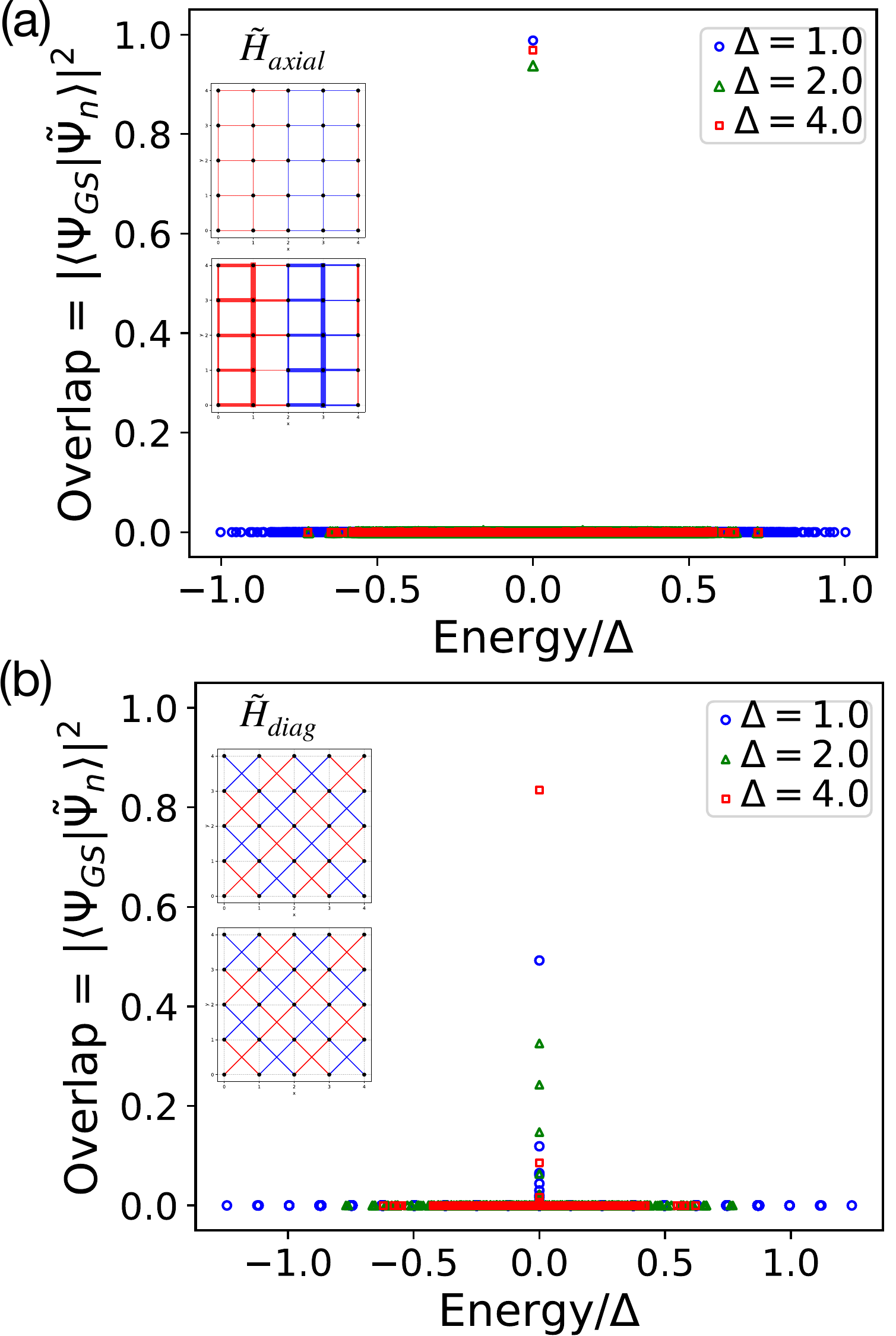}
}
\caption{\textbf{Robustness of the EHC mapping.}
Study of the overlap of the ground state of parent Hamiltonian with all the eigenstates of the EHC built target Hamiltonians computed with exact diagonalization method. (a) For \(\tilde{H}_{\mathrm{axial}}\), the overlap is negligible for most eigenstates except for a single eigenstate located in the middle of the spectrum for which the overlap approaches unity. This is independent of the value of $\Delta$. (b) 
For \(\tilde{H}_{\mathrm{diag}}\), the overlap is negligible for almost all eigenstates except for few eigenstates located in the middle of the spectrum. With increasing $\Delta$, the maximum overlap approaches unity.
}  
\label{fig:2}
\end{figure} 

\begin{figure*}[t]
\centerline{
\includegraphics[width=0.95\linewidth]{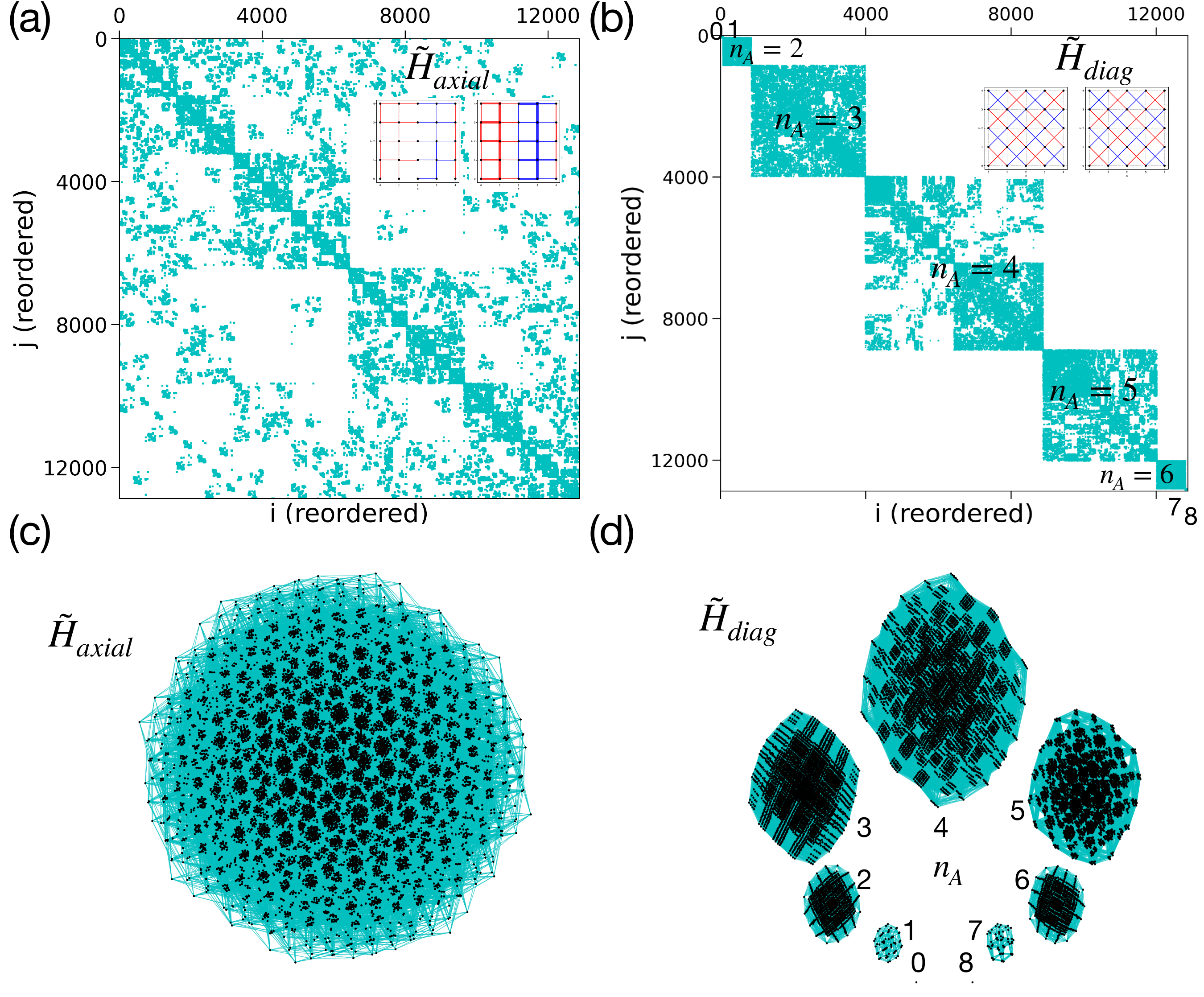}
}
\caption{\textbf{Matrix structure and Hilbert-space connectivity of target Hamiltonians.}
The figure illustrates the structure of the adjacency matrix and network representations of the EHC-engineered Hamiltonians \(\tilde{H}_{\mathrm{axial}}\), and \(\tilde{H}_{\mathrm{diag}}\). This is independent of the value of the anisotropy parameter \(\Delta\).
(a)~Sparse matrix representation of the adjacency matrix of \(\tilde{H}_{\mathrm{axial}}\) in the reordered basis obtained from the Hilbert-space connectivity graph. The Hamiltonian forms a single block, indicating that all basis states are mutually connected. 
(b)~Sparse matrix representation of the adjacency matrix of \(\tilde{H}_{\mathrm{diag}}\) in the reordered basis, which decomposes into nine disconnected blocks, corresponding to $n_A=0,\dots,8$ (see text for details). 
(c)~The network representation of \(\tilde{H}_{\mathrm{axial}}\), where nodes denote many-body basis states (represented as black dots) and edges represent nonzero off-diagonal matrix elements (represented as cyan links), revealing a single connected Hilbert-space graph.
(d)~The network representation of \(\tilde{H}_{\mathrm{diag}}\), showing multiple disconnected subgraphs that correspond to dynamically isolated sectors.
} 
\label{fig:3}
\end{figure*} 

\subsection{Robustness of the target Hamiltonians}

Using the EHC, we generate target Hamiltonians in which the ground state of the parent Hamiltonian, $|\Psi_{\mathrm{GS}}\rangle$, is deliberately promoted to an excited state. The central question for assessing the robustness of this approach is whether the mapped state corresponds to - a single excited eigenstate or a superposition of eigenstates of the target Hamiltonian. Establishing this distinction is essential, since only in the former case can the EHC procedure be regarded as yielding a well-defined result~\cite{Swain_EHC_QMC_PRB_2025}.

To investigate this, we evaluate the overlap, $\text{Overlap} = |\langle\Psi_{GS}|\tilde{\Psi}_{n}\rangle|^2 $  between the ground state of the parent Hamiltonian $|\Psi_{\mathrm{GS}}\rangle$ and each eigenstate, \(|\tilde{\Psi}_{n}\rangle\) of the target Hamiltonians $-$ $ \tilde{H}_{\mathrm{axial}}$ and $\tilde{H}_{\mathrm{diag}}$. These calculations are performed using the exact diagonalization (ED) method implemented with the QuSpin package~\cite{Quspin_2017}, which provides full access to the eigenspectrum of both models. 

\begin{figure*}[t]
\centerline{
\includegraphics[width=0.98\linewidth]{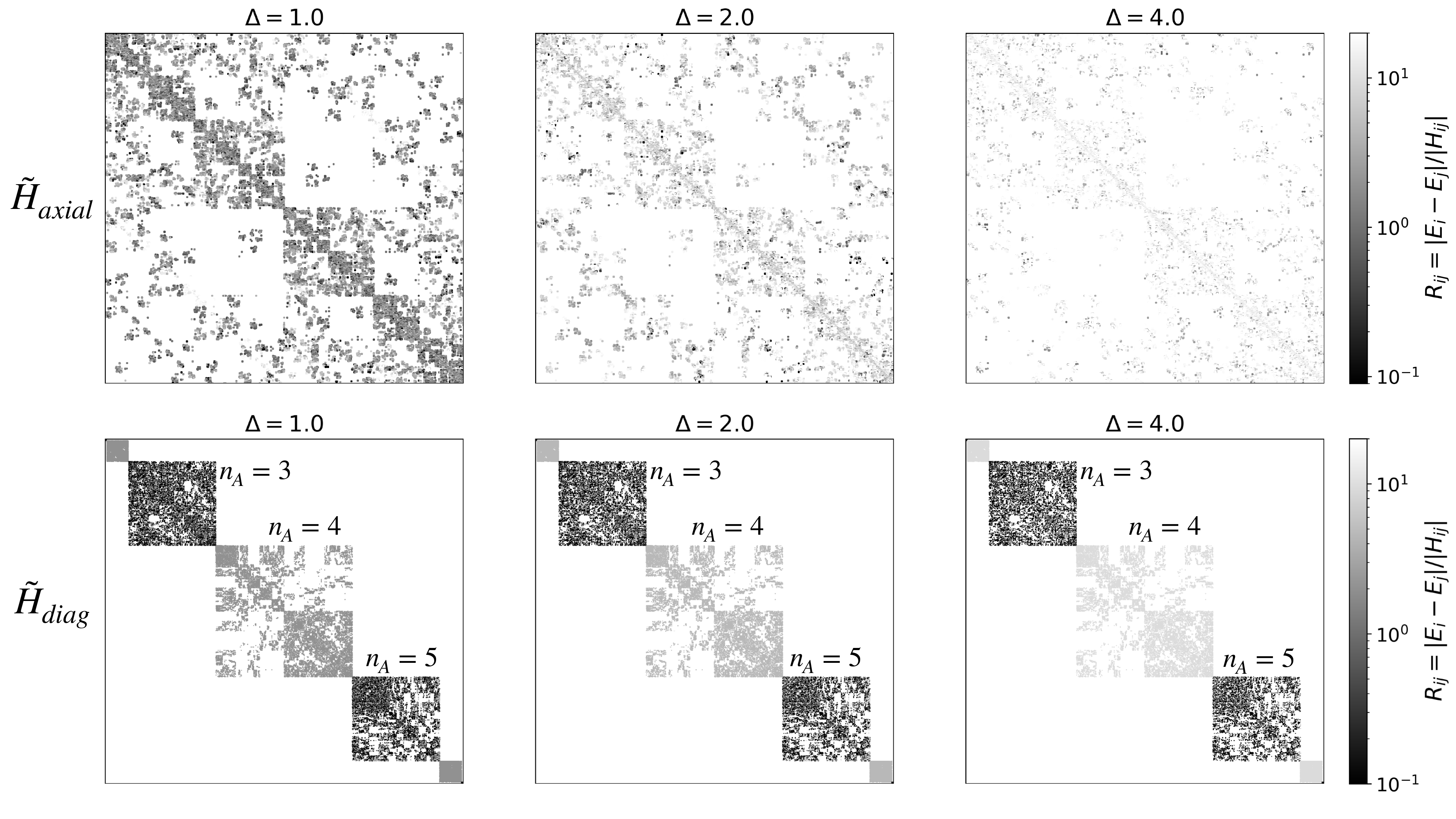}
}
\caption{
\textbf{Anisotropy-driven evolution of resonance structure for target Hamiltonians.}
Resonance matrix structure of the EHC-engineered Hamiltonians \( \tilde{H}_{\mathrm{axial}} \) and \( \tilde{H}_{\mathrm{diag}} \) with increasing anisotropy $\Delta$.  The matrix elements are characterized by the dimensionless resonance parameter \(R_{ij}=|E_i-E_j|/|H_{ij}|\), with the gray color scale indicating the degree of resonance, \(R_{ij}<1\) indicates resonant (dark) and \(R_{ij}\geq 1\) indicates off-resonant (light).
\(\tilde{H}_{\mathrm{axial}}\) forms a single connected block with a clear evolution with \(\Delta\). At large anisotropy, \(\Delta=4\), resonant links are sparse, indicating suppressed hybridization and non-ergodic tendencies.  In contrast, \(\tilde{H}_{\mathrm{diag}}\) displays a fragmented structure for all  \(\Delta\). The resonance pattern is highly inhomogeneous, with different disconnected blocks exhibiting varying densities of resonant links.  Although reducing \(\Delta\) increases the number of resonant elements within individual fragments, the absence of inter-block connectivity prevents global proliferation resulting in coexistence of ergodic and non-ergodic behavior in different blocks.} 
\label{fig:resonance}
\end{figure*} 

Fig.~\ref{fig:2}(a), shows the overlap values across the spectrum of $\tilde{H}_{\mathrm{axial}}$ for different values of the anisotropy parameter $\Delta$. It reveals that $|\Psi_{\mathrm{GS}}\rangle$ has negligible overlap with almost all eigenstates of $\tilde{H}_{\mathrm{axial}}$, except for a single eigenstate located in the middle of the spectrum. The overlap magnitude for this mid-spectrum state approches unity. This indicates that the EHC procedure produces an almost exact mapping between the parent Hamiltonian ground state and a single excited eigenstate of $\tilde{H}_{\mathrm{axial}}$. Moreover, we observe that $\Delta$ has negligible effect in this case.

In contrast, Fig.~\ref{fig:2}(b) displays the corresponding overlap spectra for $\tilde{H}_{\mathrm{diag}}$ for different values of $\Delta$. Here, the parent Hamiltonian ground state has vanishing overlap with most eigenstates, but exhibits finite overlap values for several eigenstates near the center of the spectrum. We observe that $\Delta$ has significant effect in this case. At large anisotropy, the maximum overlap approaches unity, indicating a nearly perfect one-to-one correspondence between the parent ground state and a single mid-spectrum eigenstate of the target Hamiltonian $\tilde{H}_{\mathrm{diag}}$. These results highlight the crucial role of XXZ anisotropy in stabilizing the EHC mapping and improving the reliability of the constructed target Hamiltonians.

\subsection{Structure of the target Hamiltonians}

A novel perspective on MBL is obtained by recasting the problem as an effective Anderson localization phenomenon on a complex network defined in Hilbert space~\cite{Basko2006,Monthus2010,Luitz2015,Logan2019,Tarzia2020,Tikhonov2021,Herre2023,Roy_Logan_FS_dynamics_JPCM2025}.
In this picture, many-body basis states form the nodes of a high-dimensional graph, while Hamiltonian matrix elements generate a sparse but intricate pattern of links, allowing one to characterize ergodicity breaking in terms of participation entropies and fractal dimensions~\cite{multifractal_scaling_Laflorencie,Pietracaprina2019}. Such a framework has led to significant insights into the nature of the ETH–MBL transition for the disordered S=1/2 XXZ chain in one-dimensional systems~\cite{Monthus2010,Luitz2015,Logan2019,multifractal_scaling_Laflorencie,Tarzia2020,KT_MBL_Gabriel2020,Roy_Logan_FS_dynamics_JPCM2025}.
Motivated by this, it is natural to extend this network-based analysis to more general interacting Hamiltonians, such as those constructed via the EHC approach, where nontrivial, engineered structures in Hilbert space may give rise to unconventional ergodicity breaking.

Our EHC-engineered Hamiltonians, \( \tilde{H}_{\mathrm{axial}} \) and \( \tilde{H}_{\mathrm{diag}} \) are highly sparse matrices in the many-body basis, with the sparsity pattern determined by the underlying interaction terms (see fig.\ref{fig:1}). To probe whether these Hamiltonians possess an intrinsic block diagonal structure, we reinterpret them as networks defined over the Hilbert space. In this representation, each many-body basis state is denoted as a node, and any nonzero off-diagonal matrix element of the Hamiltonian defines a link between the corresponding nodes. The adjacency matrix captures the connectivity structure of the Hilbert space independent of the basis ordering~\cite{newman2006structure,hagberg2008exploring}.

From a network perspective, if every node can be reached from every other through these links, the network is fully connected~\cite{newman2006structure}. If instead the nodes split into groups that are not mutually connected, the network breaks into disconnected clusters. By reordering the basis so that states within the same cluster are grouped together, the Hamiltonian matrix can be brought into a block-diagonal form, with each block corresponding to one cluster. This provides a simple way to determine whether the Hilbert space is fully connected or fragmented into independent sectors. 
This construction is mainly based on off-diagonal connectivity. The diagonal matrix elements, which correspond to the on-site energies in the network picture, can also be consistently retained providing energy relevant information.

We show the results of applying this procedure to our EHC-engineered Hamiltonians in Fig.~\ref{fig:3}.
For \( \tilde{H}_{\mathrm{axial}} \), we find that the reordered Hamiltonian consists of a single block only, indicating that all basis states are mutually connected through the action of the Hamiltonian (see panel (a)). In network terms, the Hilbert space forms a single, densely connected graph, consistent with the absence of Hilbert-space fragmentation (panel (c)).

The same analysis applied to \( \tilde{H}_{\mathrm{diag}} \), shows nine disconnected blocks (see panel (b)).  
The origin of this can be understood as follows. For our $4\times 4$ lattice with periodic boundaries, the interactions along NNN bonds act independently within the two checkerboard sublattices $A$ and $B$. Consequently, the dynamics conserve the sublattice magnetizations separately, and the system develops an additional conserved quantum number $n_A$, defined as the number of up spins on sublattice $A$. In the $S_{\mathrm{tot}}^z=0$ sector, this restricts $n_A$ to values $0,\dots,8$, leading to a decomposition of the Hilbert space into dynamically disconnected sectors labeled by $n_A$. 
Furthermore, for a fixed $n_A$, the number of configurations factorizes between the two sublattices, giving a block dimension $\mathcal{D}(n_A)=\binom{8}{n_A}\binom{8}{8-n_A}=\left[\binom{8}{n_A}\right]^2$. 
This produces nine sectors with dimensions \{1,64,784,3136,4900,3136,784,64,1\}, corresponding to $n_A=0,\dots,8$. The largest block occurs for $n_A=4$.
Fig.~\ref{fig:3}(d) reveals the Hilbert space network of \( \tilde{H}_{\mathrm{diag}} \)
corresponding to $n_A=0,\dots,8$. 
The stark contrast between the two cases indicates how the EHC approach enables engineering of target Hamiltonians with different Hilbert-space connectivity.

\begin{figure*}[t]
\centerline{
\includegraphics[width=0.98\linewidth]{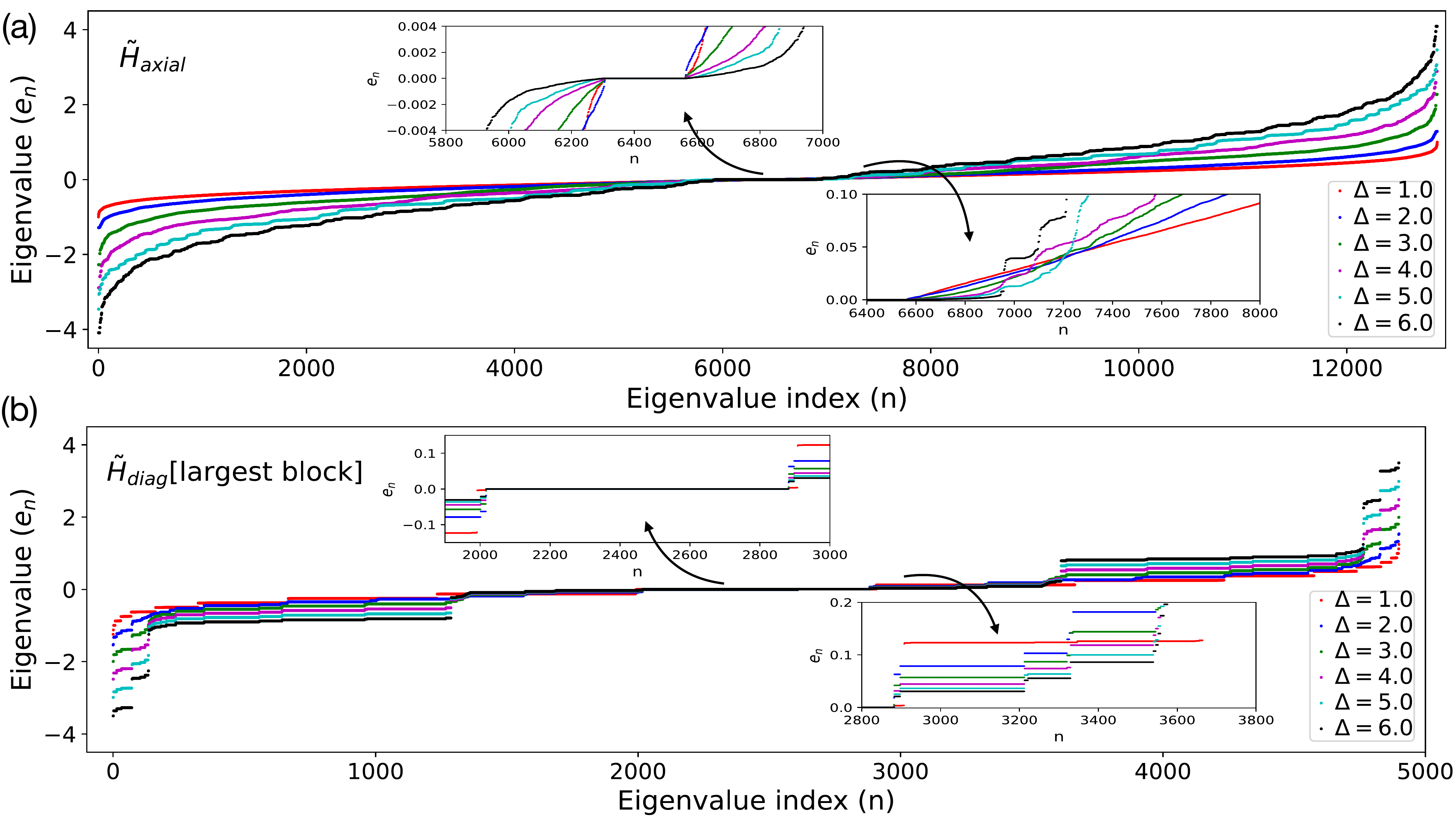}
}
\caption{\textbf{Eigenspectra of target Hamiltonians:}
Full many-body eigenspectra of (a) \( \tilde{H}_{\mathrm{axial}} \) and (b) \( \tilde{H}_{\mathrm{diag}} \) for different anisotropy \(\Delta\).
Both models exhibit an extensive zero-energy degeneracy at mid-spectrum (left insets), robust across anisotropy values.
For \( \tilde{H}_{\mathrm{axial}} \), the spectrum evolves from quasi-continuous band edges at \(\Delta=1-3\) to a staircase-like structure at larger \(\Delta\) values (right inset).
In contrast, the largest block of \( \tilde{H}_{\mathrm{diag}} \) displays a discrete, strongly step-like spectrum for all anisotropy values with decreasing mid-spectrum degeneracy as anisotropy increases (see insets).
} 
\label{fig:5}
\end{figure*} 

\subsection{Resonances in the target Hamiltonians}

The sparsity pattern and graph connectivity of the EHC-engineered target Hamiltonians shown in fig~\ref{fig:3} are independent of the anisotropy parameter $\Delta$. However, their dynamical character is dictated by how local configurations hybridize in Hilbert space, which is tuned by $\Delta$. Inspired by the studies of resonance criterion in disordered systems~\cite{Altshuler1997,Basko2007,Gornyi2005,DeRoeck2017,Thiery2018,Luitz2017,Tarzia2024,Tarzia2025}, we employ a similar resonance criterion defined directly for our Hamiltonian matrix. For two basis states $i$ and $j$, connected by a nonzero off-diagonal matrix element $H_{ij}$ we define the dimensionless ratio $R_{ij}=|E_i-E_j|/|H_{ij}|$, where $E_i$ and $E_j$ are the corresponding diagonal (on-site) energies. 
Matrix elements with $R_{ij} < 1$ are identified as resonant, indicating that the coupling dominates over the energy mismatch and enables efficient hybridization, while elements with $R_{ij}\geq 1$ corresponds to off-resonant links. This criterion provides a transparent probe of how the ratio of energy mismatch and hybridization translates into dynamical behavior. We expect a dense structure of resonant links to promote ergodicity via delocalization in the Hilbert space, while a sparse or isolated resonances will favor non-ergodic dynamics~\cite{DeRoeck2017,Thiery2018,Luitz2017,Morningstar2022,Ha2023ResonancesAvalanche}. For visualization we plot $R_{ij}$ on a color scale, and study its evolution with varying $\Delta$ (see fig.~\ref{fig:resonance})

Applying this framework to our models uncovers sharply distinct anisotropy-dependent behaviors (see fig~\ref{fig:resonance}). 
For \(\tilde{H}_{\mathrm{axial}}\), whose Hilbert-space graph forms a single connected component, the resonance structure undergoes a clear crossover with decreasing $\Delta$. At large anisotropy \(\Delta=4\), resonant links are sparse, suppressing hybridization despite the high connectivity of the graph and leading to non-ergodic features. As \(\Delta\) is reduced, the density of resonant elements increases. This proliferation of resonances enables crossover to ergodic behavior.

In contrast, \(\tilde{H}_{\mathrm{diag}}\) exhibits an intrinsically fragmented block structure, and the resonance analysis reveals a more heterogeneous feature~\cite{HSF_Sala_PRX2020,HSF_local_PRB_2020,scar_rev_Moudgalya_2022}. As seen in fig.~\ref{fig:resonance}, at \(\Delta=4\), different disconnected blocks display markedly distinct resonance densities, some blocks host a significant large number of resonant links, while others remain largely off-resonant. This inhomogeneous resonance structure naturally leads to coexistence of dynamical behaviors across fragments, underpinning the mixture of ergodic and non-ergodic eigenstates observed in the spectrum (shown later). On reducing \(\Delta\), resonances become more prevalent within individual fragments, yet remain confined by the underlying fragmentation, preventing global proliferation. The interplay between fragmentation and resonance proliferation provides a microscopic picture for the emergence of non-ergodic dynamics in our EHC-engineered Hamiltonians, setting the stage for the spectral and dynamical analyses that follows next.

\begin{figure*}[t]
\centerline{
\includegraphics[width=0.99\linewidth]{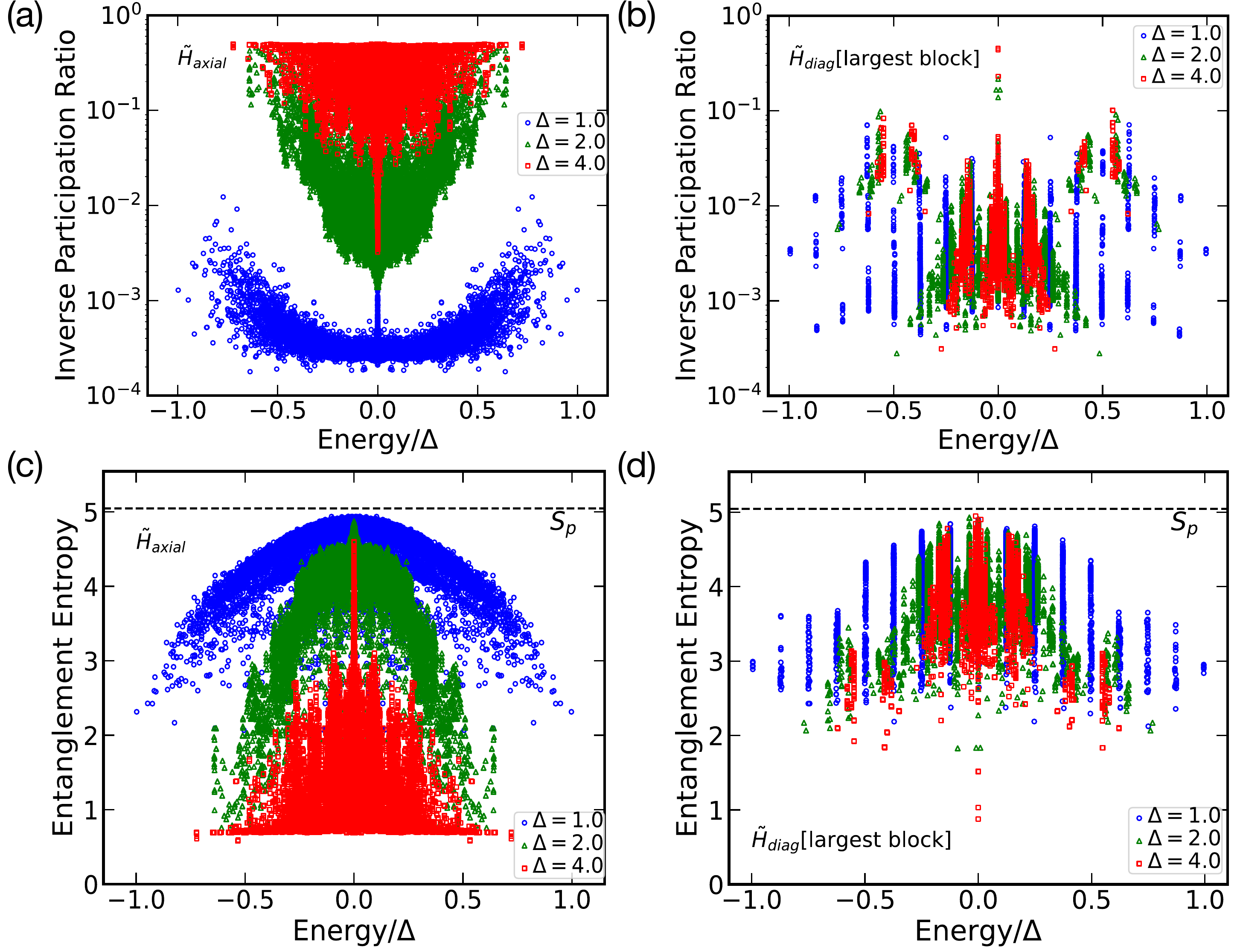}
}
\caption{\textbf{Spectral diagnostics of ergodicity breaking: IPR and entanglement entropy.}
Energy-resolved inverse participation ratio (IPR) and von Neumann entanglement entropy (EE) of the target Hamiltonians - \(\tilde{H}_{\mathrm{axial}}\) and \(\tilde{H}_{\mathrm{diag}}\) for varying  $\Delta$.  For \(\tilde{H}_{\mathrm{axial}}\), both the IPR and EE spectra reveal a clear anisotropy-driven crossover. At small $\Delta$, dense resonant hybridization produces low IPR and volume-law entanglement consistent with ergodic behavior. Increasing $\Delta$ suppresses resonances as seen in Fig.~\ref{fig:resonance}, leading to the emergence of high-IPR, low-EE eigenstates and ultimately a strongly non-ergodic regime.
In contrast, the largest block of \(\tilde{H}_{\mathrm{diag}}\) exhibits pronounced tower-like structures in both IPR and EE for all $\Delta$ values. With increasing $\Delta$, these towers broaden in a gradual manner while retaining approximate energy regularity. The resulting high-IPR, low-EE states embedded within the ergodic background constitute quantum many-body scar states.}
\label{fig:6}
\end{figure*} 

A natural question to address is whether the apparent nonergodic features of \(\tilde{H}_{diag}\) originate simply from the fragmentation of the Hilbert space or from more intrinsic properties of the EHC-built Hamiltonian.
To understand this, it is essential to compare the eigenstate properties of \(\tilde{H}_{diag}\) with \(\tilde{H}_{axial}\), whose Hilbert space does not possess any fragmentation. Such a comparison allows us to isolate the role of fragmentation from other mechanisms that may influence the eigenstate structure, spectral properties, and dynamics.

Within \(\tilde{H}_{diag}\), the most meaningful comparison is obtained by focusing on its largest irreducible block. 
This sector contains the overwhelming majority of the Hilbert-space states and therefore captures the typical many-body behavior of the fragmented model, while avoiding finite-size artifacts associated with the much smaller sectors. 
Moreover, each disconnected block evolves independently under the dynamics, implying that nonergodic properties such as inverse participation ratio, entanglement entropy and wave-packet dynamics etc. must be assessed within individual blocks rather than in the full fragmented Hilbert space.
Therefore, studying the largest block enables a direct investigation of whether the nonergodic phenomena observed in \(\tilde{H}_{diag}\) persist even after the trivial effects of fragmentation are removed.

For this reason, in this work we compare the spectral and dynamical properties of \(\tilde{H}_{axial}\) with those of the largest irreducible sector of \(\tilde{H}_{diag}\). 

\subsection{Eigenspectra of the target Hamiltonians}

We present the many-body eigenspectra of the EHC-engineered Hamiltonians \(\tilde{H}_{\mathrm{axial}}\) and the largest block of \(\tilde{H}_{\mathrm{diag}}\) as a function of the anisotropy parameter in Fig.~\ref{fig:5}. 
A prominent and robust feature common to both models is the appearance of an extensive degeneracy at the center of the spectrum, pinned at zero energy.
As highlighted in the left insets of panels (a) and (b), this mid-spectrum degeneracy persists across the entire range of \(\Delta=1-6\). This seems to be a structural property inherited from the EHC construction rather than fine-tuning of parameter choices. Furthermore, the eigenspectra are reminiscent of the devil's staircase seen e.g.\  an one-dimensional Ising model with long-range antiferromagnetic interactions~\cite{PhysRevLett.49.249}. 

Beyond this shared feature, the global structure of the spectrum differs markedly between the two models. 
For \(\tilde{H}_{\mathrm{axial}}\) the spectrum near the band edges is continuous for weak to intermediate anisotropy, \(\Delta=1-3\), indicative of a densely connected Hilbert space compatible with thermal behavior. With increasing \(\Delta\), clear gaps develop between successive energy levels, and the eigenspectrum acquires a staircase-like structure, as shown in the right inset of Fig~\ref{fig:5}(a). 
This suggests a crossover behavior of the energy eigenspectra between weaker anisotropy (\(\Delta\sim 1\)) and stronger anisotropy (\(\Delta\sim 4\)). 
In contrast, the eigenspectrum of the largest block of \(\tilde{H}_{\mathrm{diag}}\) is discrete for all anisotropy values, \(\Delta=1-6\). The eigenspectrum exhibits pronounced jumps and a staircase-like form (see right inset of Fig~\ref{fig:5}(b)).
The zero-energy degeneracy remains for all \(\Delta\) values. 
The distinct spectral features suggest that eigenstates of \(\tilde{H}_{\mathrm{axial}}\) and \(\tilde{H}_{\mathrm{diag}}\) may exhibit different non-ergodic properties, which we probe next.

\subsection{Spectral properties of the target Hamiltonians}

{\it Inverse Participation Ratio:} To further probe the nature of the eigenstates, we compute the inverse participation ratio (IPR) of the eigenstates in the computational basis ${\rm IPR}_n = \sum_{\alpha=1}^{\mathcal N} |\langle \alpha | \psi_n \rangle|^4$,
where $\{ |\alpha\rangle \}_{\alpha=1}^{\mathcal N}$ is the many-body computational basis and $\{ |\psi_n\rangle\}_{n=1}^{\mathcal N}$ are the the eigenstates of our target Hamiltonian.
${\rm IPR}_n$ is a measure of the inverse volume occupied by an eigenstate in Hilbert space and it quantifies the extent of delocalization of a state in the Hilbert space~\cite{Evers2008,Luitz2014,multifractal_scaling_Laflorencie}. Ergodic eigenstates typically spread over an exponentially large number of the Hilbert space yielding small IPR values ($\mathrm{IPR}\sim 1/\mathcal N$)~\cite{Luitz2014,multifractal_scaling_Laflorencie} In contrast, non-ergodic or scar states are localized over a small subset of the Hilbert space, producing anomalously large IPR values ($\mathrm{IPR}\gg 1/\mathcal N$). Thus, the IPR spectrum serves as a sensitive probe of breaking of ergodicity and the presence of scar eigenstates.

{\it Axial model:} In Fig.~\ref{fig:6}(a), we show the IPR spectrum of $\tilde{H}_{\mathrm{axial}}$ as a function of energy for increasing anisotropy $\Delta$. At $\Delta=1$, the spectrum is dominated by low IPR values (IPR $\sim 1/\mathcal N$) consistent with fully ergodic eigenstates satisfying ETH, apart from the trivial degeneracy at the band center. This behavior can be understood from our resonance analysis, where the resonance matrix supports a dense and increasing number of resonant links $(R_{ij}<1)$, enabling efficient hybridization across the entire matrix and promoting ergodic dynamics.
At intermediate anisotropy, $\Delta=2$, a subset of eigenstates with enhanced IPR emerges in the mid-spectrum, coexisting with an otherwise thermal background. This signals the onset of suppression of resonances. While the resonance matrix remains globally connected, the density of resonant links decreases, leading the appearance of non-ergodic eigenstates.
At large anisotropy, $\Delta=4$, this trend becomes more pronounced. The resonance network is sparse and fails to percolate effectively despite the underlying connectivity of the matrix. As a consequence, the IPR spectrum exhibits a broad distribution with many eigenstates having IPR $\gg 1/\mathcal N$, including a substantial fraction with IPR $\approx 0.5$, indicative of strong localization over close to two many-body basis states, a feature of strong non-ergodic behavior.   
This establishes a clear, anisotropy-driven crossover from ergodic to non-ergodic behavior, governed by the progressive suppression of resonant hybridization in the Hilbert space.

{\it Diagonal model:} The IPR spectrum corresponding to the largest irreducible block of $\tilde{H}_{\mathrm{diag}}$, shown in Fig.~\ref{fig:6}(b), reveals a very different structure. Even at $\Delta=1$, the spectrum lacks a uniform thermal background, and instead exhibits sharply defined tower-like structures. These towers contain degenerate eigenstates spanning a wide range of IPR values, from ergodic ($IPR \sim 1/\mathcal{N}$) to strongly non-ergodic ($IPR \gg 1/\mathcal{N}$), at the same energy. 
This behavior reflects the underlying resonance landscape of the largest block, where resonance amplitudes are weak. Strong resonance amplitudes in other blocks of $\tilde{H}_{\mathrm{diag}}$ remain confined to those blocks only, thus being unable to induce global ergodic dynamics~\cite{HSF_Sala_PRX2020,HSF_local_PRB_2020,scar_rev_Moudgalya_2022}. 
As $\Delta$ increases, the tower structures persist, but undergo systematic broadening, thus a reduction in degeneracy. 
The approximate energy regularity of these towers still remains. We interpret these structured, high-IPR eigenstates as quantum many-body scar states of our model~\cite{Khemani_scar_XY_2020}.

\begin{figure*}[t]
\centerline{
\includegraphics[width=0.85\linewidth]{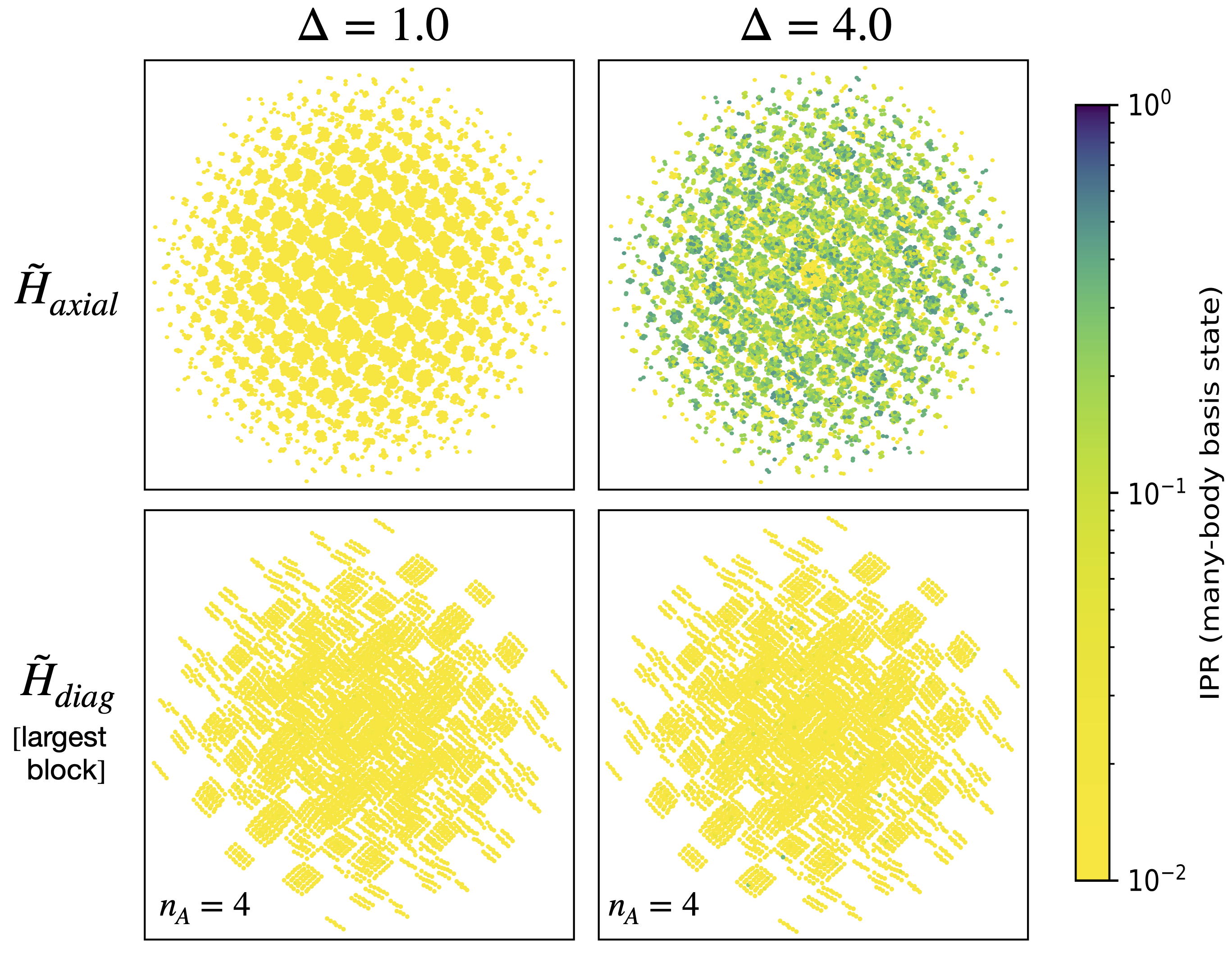}
}
\caption{
\textbf{Hilbert-space representation of target Hamiltonians for different anisotropy values.}
Each node represents a many-body basis state, 
with node colors encoding ${\rm IPR}_{\alpha}$. 
For \(\tilde{H}_{\mathrm{axial}}\), the basis states exhibit uniformly low IPR at \(\Delta=1\), while increasing anisotropy to \(\Delta=4\) produces states with enhanced IPR values.
For the largest block of \(\tilde{H}_{\mathrm{diag}}\),
all basis states display comparable, yet elevated IPR values relative to \(IPR \sim 1/\mathcal N\), even at \(\Delta=1\). Increasing the anisotropy to \(\Delta=4\) further enhances this effect, resulting in some basis states with distinctly large IPR values.
} 
\label{fig:7}
\end{figure*} 

{\it Entanglement Entropy (EE):} An effective way to distinguish non-ergodic eigenstates from ergodic ones is by studying the scaling behavior of entanglement entropy (EE) at finite energy density~\cite{PhysRevE.86.010102,PhysRevE.86.010102,DAlessio2016}. In ergodic systems, typical high energy excited states follow ETH and exhibit volume-law scaling, with the von-Neumann entanglement entropy of a subsystem A scaling as $S_\mathrm{E} \propto V_A$.
In contrast, non-ergodic eigenstates display markedly reduced entanglement growth, and often consistent with logarithmic or otherwise sub-volume scaling~\cite{Abanin2019,scar_2018_Moudgalya}. This sub-volume scaling of entanglement is a sign of ergodicity breaking.
To quantify further, we compute the von-Neumann entanglement entropy, $S_\mathrm{E}=-\mathrm{Tr}\rho_\mathrm{A}\log\rho_\mathrm{A}$, where $\rho_\mathrm{A}$ is the reduced density matrix of subsystem A. 
Using exact diagonalization method, and choosing the left-half of our lattice as the subsystem A, we compute $S_\mathrm{E}$ across the entire spectrum of the target Hamiltonian, $ \tilde{H}_{\mathrm{axial}}$ and $\tilde{H}_{\mathrm{diag}}$. The results are shown in Fig.~\ref{fig:6}(c-d). 

{\it Axial model:} The EE spectrum of $\tilde{H}_{\mathrm{axial}}$, shown in Fig.~\ref{fig:6}(c), provides a complementary perspective on the ergodic to non-ergodic crossover inferred from the IPR analysis. 
For $\Delta=1$, the EE spectrum exhibits a broad distribution with most eigenstates attaining values close to the thermal value (bounded by Page entropy, $S_P$). This is consistent with volume-law entanglement and ETH behavior. This is also in agreement with the resonance structure. The resonance matrix supports a dense set of resonant links, enabling extensive hybridization and resulting in highly entangled eigenstates.
At intermediate anisotropy, $\Delta=2$, while a majority of eigenstates continue to display near-thermal entanglement, a distinct subset of mid-spectrum states emerges with significantly reduced EE compared to their surroundings. This signals the onset of non-ergodic structure and reflects the partial suppression of resonances observed in the resonance matrix. This gives rise to a coexistence of strongly hybridized (thermal) and weakly hybridized (non-ergodic) eigenstates.
For $\Delta=4$, the EE spectrum undergoes a dramatic restructuring $-$ the entanglement of all eigenstates is substantially reduced relative to the thermal bound. This behavior mirrors the near-complete suppression of resonant links in the resonance matrix (see fig.~\ref{fig:resonance}), which inhibits hybridization across configurations despite the underlying connectivity. The resulting eigenstates exhibit low entanglement and strong localization in Hilbert-space, establishing a robust non-ergodic regime. The EE spectrum reinforces the picture of an anisotropy-driven crossover, where ergodicity is progressively destabilized by the suppression of resonant processes. 

{\it Diagonal model:} The EE spectrum corresponding to the largest irreducible block of $\tilde{H}_{\mathrm{diag}}$, reveals a qualitatively distinct structure.
For $\Delta=1$, the EE spectrum completely deviates from a uniform thermal profile and instead displays sharply defined, highly degenerate tower-like features. 
These towers contain eigenstates with anomalously low entanglement entropy, coexisting with more thermal states at similar energies. 
As $\Delta$ increases, these EE-towers persist, but broaden gradually and lose degeneracy, mirroring the evolution observed in the IPR spectrum. 
Despite this broadening, the tower-like organization and approximate energy regularity remains intact, indicating the persistence of structured, weakly entangled eigenstates.
The coexistence of these low-EE tower states with a background of more entangled states provides a clear signature of weak ergodicity breaking. These atypical eigenstates, characterized by both low entanglement and high IPR, have characteristics of quantum many-body scar states. The EE spectrum thus corroborates the unified picture emerging from our resonance and IPR analyses.

\begin{figure*}[t]
\centerline{
\includegraphics[width=0.98\linewidth]{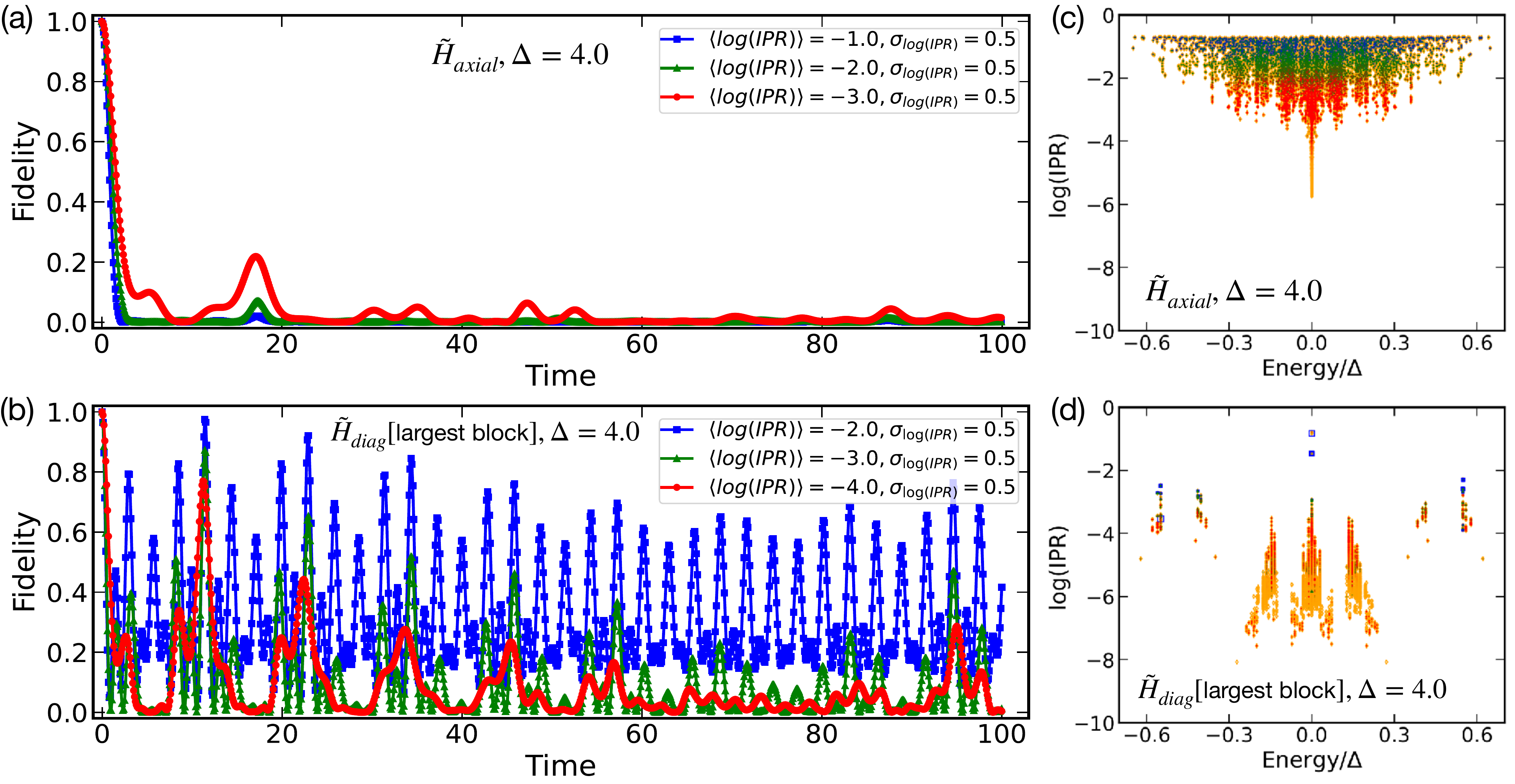}
}
\caption{\textbf{Fidelity revivals and eigenstate overlap structure in target Hamiltonians:}
(a)–(b)~Comparison of the time evolution of the fidelity, $F(t)=|\langle \Psi_{in}(0)|\Psi_{in}(t)\rangle|^2$, or an initial Gaussian wave-packet state parametrized by $\langle \log(\mathrm{IPR}) \rangle$ and fixed $\sigma_{\log(\mathrm{IPR})} = 0.5$ under the target Hamiltonians for $\Delta=4.0$.  
For \(\tilde{H}_{\mathrm{axial}}\), with the wave-packet parametrized by $\langle \log(\mathrm{IPR}) \rangle = -1.0$, the fidelity decays rapidly and saturates to a negligible value at long times, indicating effective thermalization and loss of memory of the initial state. Shifting the wave-packet center towards lower $\langle \log(\mathrm{IPR}) \rangle$ increases the number of participating eigenstates, further enhancing dephasing and reinforcing ergodic relaxation.
In contrast, for the largest block of \(\tilde{H}_{\mathrm{diag}}\), the fidelity exhibits pronounced long-lived revivals for wave-packet with $\langle \log(\mathrm{IPR}) \rangle = -2.0$, signaling non-ergodic dynamics associated with scar eigenstates. As the wave-packet center is shifted towards lower $\langle \log(\mathrm{IPR}) \rangle$, the revival amplitude diminishes and eventually disappears. 
(c)–(d)~Overlap distribution $|\langle \Psi_{in} | \tilde{\Psi}_n \rangle|^2$ superimposed on the IPR spectrum. The logarithm of the IPR is shown in orange color data, while the overlap amplitudes are represented by symbol sizes proportional to the magnitude, and with colors corresponding to different wave-packet parameters as indicated in (a)–(b). For \(\tilde{H}_{\mathrm{axial}}\), with the wave-packet parametrized by $\langle \log(\mathrm{IPR}) \rangle = -1.0$, the overlap is broadly distributed over a (quasi) continuum of eigenstates, leading to a wide frequency spectrum and the absence of revivals. Shifting the wave-packet center further broadens this distribution, explaining the rapid dephasing and absence of revivals.
In contrast, for the largest block of \(\tilde{H}_{\mathrm{diag}}\), the overlap is sharply concentrated on the discrete high-IPR tower states with near-regular energy spacing, resulting in a small set of dominant frequencies and coherent revival dynamics for wave-packet with $\langle \log(\mathrm{IPR}) \rangle = -2.0$. As the wave-packet center is shifted to lower $\langle \log(\mathrm{IPR}) \rangle$, the overlap spreads over multiple frequencies, leading to suppression of revivals.}
\label{fig:8}
\end{figure*} 

A complementary perspective is obtained by studying the IPR of a computational basis state expressed in the eigenbasis,
${\rm IPR}_{\alpha} = \sum_{n=1}^{\mathcal N} |\langle \psi_n | \alpha \rangle|^4$, which investigates how individual basis configurations spread over the eigenstates~\cite{PhysRevB.100.214313,Roy_Logan_FS_dynamics_JPCM2025}. The Hilbert-space representation of the EHC-engineered target Hamiltonians are shown in Fig.~\ref{fig:7}. In this representation, many-body basis states are treated as nodes of a Hilbert-space graph, with ${\rm IPR}_{\alpha}$ encoding the color of each node.
This construction directly links spectral diagnostics to the underlying Hilbert-space geometry, allowing us to determine how nonergodic behavior emerges in our EHC engineered models.

For \(\tilde{H}_{\mathrm{axial}}\), the Hilbert-space representation remains a single connected sector at both \(\Delta=1\) and \(4\). At \(\Delta=1\), the basis state representation is dominated by states with IPR values close to the fully ergodic expectation \(IPR \sim 1/\mathcal N\), consistent with predominantly thermal eigenstates. Upon increasing the anisotropy to \(\Delta=4\), a pronounced inhomogeneity emerges, with a subset of the basis states exhibiting substantially enhanced IPR values, \(IPR \gg 1/\mathcal N\). Thus, increased anisotropy promotes increased non-ergodic behavior within a globally connected Hilbert space. Furthermore, the Hilbert-space representation explicitly shows how non-ergodic eigenstates coexist with ergodic ones in \(\tilde{H}_{\mathrm{axial}}\).

As shown in fig.\ref{fig:3}(d), the Hilbert-space graph of \(\tilde{H}_{\mathrm{diag}}\) decomposes into multiple disconnected subgraphs. 
As a result, the IPR values of the basis states are not dictated by the full Hilbert-space size but by the size of each fragmented subspace. 
For the largest block of \(\tilde{H}_{\mathrm{diag}}\), even at \(\Delta=1\), all basis states display comparable, yet elevated IPR values relative to \(IPR \sim 1/\mathcal N\), reflecting the reduced accessible Hilbert space within individual subgraphs rather than strong localization in a conventional sense.
Increasing the anisotropy to \(\Delta=4\), further enhances this effect, resulting in several basis states with very large IPR values.

\subsection{Wave-packet dynamics in the eigenstate basis}
\label{dyn_real_time}

A hallmark of quantum many-body scar systems is their anomalous real-time dynamics when initialized in carefully prepared states with significant overlap on non-ergodic eigenstates~\cite{scar_states_Turner2018,scar_expt_driven_rydberg_2021}. In contrast to generic initial states, which rapidly thermalize under non-integrable dynamics, such states exhibit long-lived coherent oscillations and periodic revivals, reflecting the presence of a small subset of atypical eigenstates embedded within an otherwise thermal spectrum~\cite{scar_states_Turner2018_2,scars_2020_Iadecola,scar_states_Serbyn2021}. The persistence of these revivals, often surviving well beyond the timescales set by typical relaxation processes, provides a clear dynamical fingerprint of weak ergodicity breaking in scar models~\cite{scar_states_Turner2018,scar_states_Turner2018_2,scars_2020_Iadecola,scar_states_Serbyn2021,scar_expt_driven_rydberg_2021}.

To probe this behavior, we construct a family of Gaussian wave-packet states in the eigenbasis of the target Hamiltonians, guided by the distribution of inverse participation ratio (${\rm IPR}_n$) values in our study. We design an initial wave-packet state $|\Psi_{in} \rangle \equiv  \frac{1}{N}\sum_{m} c_m e^{-\frac{1}{2} \left(\frac{\log (IPR)_m - \langle \log(\mathrm{IPR}) \rangle}{\sigma_{\log(\mathrm{IPR})}}\right)^2} |\tilde{\Psi}_m\rangle$ parameterized by the mean, $\langle \log(\mathrm{IPR}) \rangle$, and the standard deviation, $\sigma_{\log(\mathrm{IPR})}$, of the logarithm of the IPR of the eigenstates $|\tilde{\Psi}_m\rangle$. $c_m \in [0,1]$ are real, uniformly distributed random numbers and $N$ is the normalization factor. $\langle \log(\mathrm{IPR}) \rangle$ sets the spectral center of the wave packet, while $\sigma_{\log(\mathrm{IPR})}$ controls its spectral spread.
These tailored wave packets are then unitarily evolved under the target Hamiltonians obtained via EHC for anisotropy parameter $\Delta=4$. We compute the fidelity, $F(t) = |\langle \Psi_{in}(0) | \Psi_{in}(t) \rangle|^2$, to probe its dynamical behavior. It quantifies the degree of state revival during unitary time evolution.

As shown in fig~\ref{fig:8}(a), for \(\tilde{H}_{\mathrm{axial}}\), the fidelity decays rapidly and saturates to a negligible value at long times, indicating efficient dephasing and loss of memory of the initial state, consistent with ergodic dynamics. In contrast, for the largest block of \(\tilde{H}_{\mathrm{diag}}\), fidelity exhibits pronounced and long-lived revivals across families of wave packets, as seen in fig.~\ref{fig:8}(b), with the strongest revivals arising when the wave packet is centered on high-IPR states. 
This behavior follows from the underlying spectral structure $-$ while \(\tilde{H}_{\mathrm{axial}}\) supports a quasi-continuum of eigenstates leading to a broad distribution of frequencies and rapid dephasing, the largest block of \(\tilde{H}_{\mathrm{diag}}\) hosts structured towers of eigenstates with approximately regular energy spacings (see fig~\ref{fig:5}). The sharp IPR towers in the spectrum of  \(\tilde{H}_{\mathrm{diag}}\) (see panel (d)) introduce a small set of dominant frequencies, enabling coherent interference and revival dynamics. As we reduce $\langle \log(\mathrm{IPR}) \rangle$ from $-2.0$ to $-4.0$, the spectral weight spreads over a larger number of eigenstates, the frequency content broadens and the revivals are progressively suppressed. This behavior indicates the presence of dynamical signatures of scar states in \(\tilde{H}_{\mathrm{diag}}\), providing evidence that our EHC approach yields target Hamiltonians with scar-like non-ergodic dynamics.

\begin{figure*}[t]
\centerline{
\includegraphics[width=0.9\linewidth]{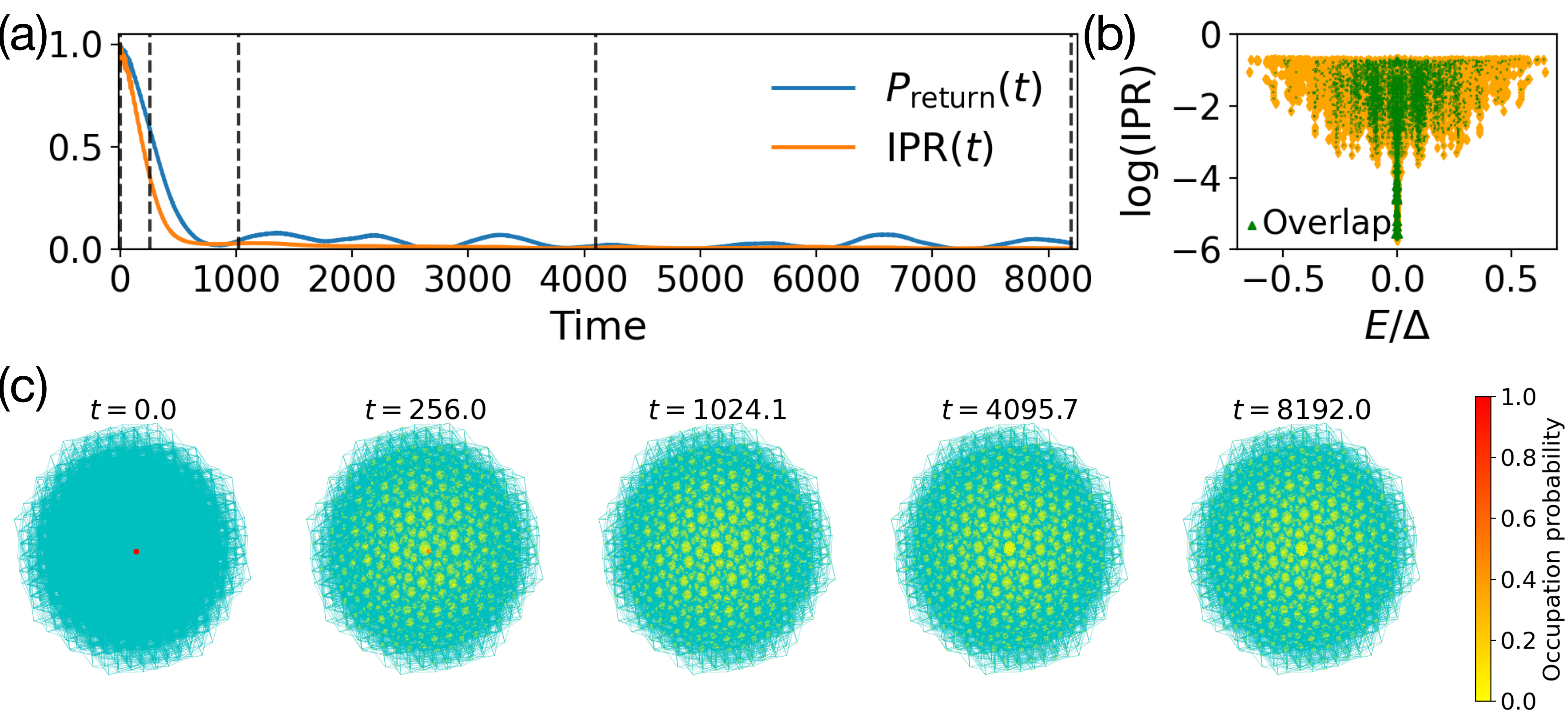}
}
\centerline{
\includegraphics[width=0.9\linewidth]{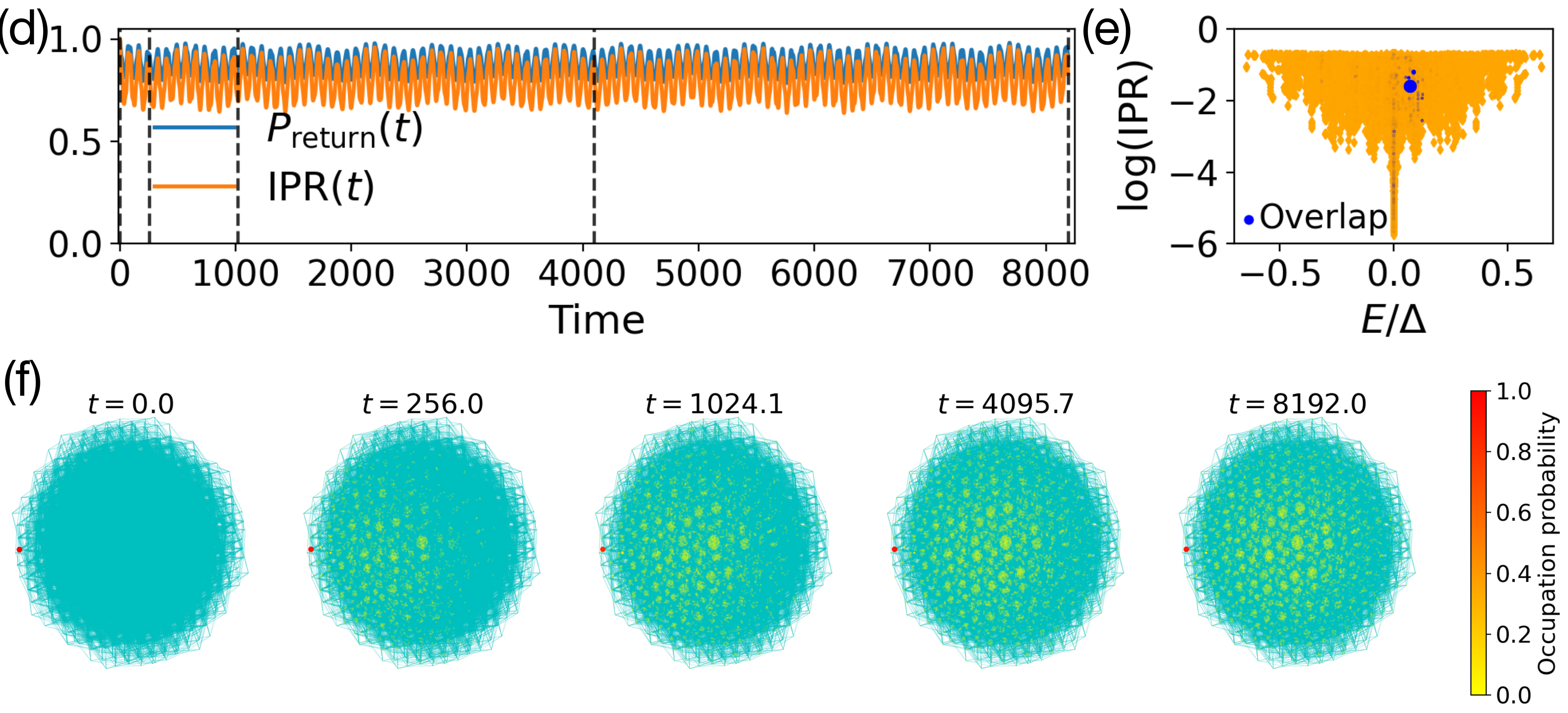}
}
\caption{
\textbf{Fock-space dynamics for \(\tilde{H}_{\mathrm{axial}}\) with \(\Delta=4\):}
Unitary time evolution of representative Fock-space basis states illustrating contrasting dynamical behavior on a fully connected Hilbert-space network. At $t=0$, the occupation probability is unity on the initial node and zero elsewhere.  The top panels (a)–(c) show a delocalized initial state ($\mathrm{IPR}_{\alpha} \approx 0.005$) with node connectivity=20, while the bottom panels (d)–(f) show a localized initial state ($\mathrm{IPR}_{alpha} \approx 0.5$) with node connectivity=12.  (a) The return probability and time-dependent IPR decay rapidly, indicating ergodic behavior.
(b) The overlap distribution $|\langle \Psi_{\alpha} | \tilde{\Psi}_n \rangle|^2$, superimposed on the IPR spectrum, is broadly spread over many eigenstates. The logarithm of the IPR is shown in orange color data, while the overlap values are represented by green symbols with sizes proportional to the overlap magnitude.
(c) Fock-space snapshots show rapid and widespread redistribution of probability across the network, consistent with ergodic dynamics.  (d) The return probability and IPR remain close to their initial values, indicating strong memory retention of initial state, characteristic of non-ergodic behavior. (e) The overlap distribution superimposed on the IPR spectrum, is sharply concentrated on a small subset of eigenstates. The logarithm of the IPR is shown in orange, while the overlap values are represented by blue symbols with sizes proportional to the overlap magnitude. (f) Fock-space snapshots reveal slow spreading of probability, with dominant weight near the initial node, signaling non-ergodic dynamics despite the underlying connectivity.}
\label{fig:9}
\end{figure*} 

\subsection{Eigenstate overlap structure and origin of fidelity revivals}

To gain a deeper understanding of the origin of the observed dynamics, we compute the overlap of our initial Gaussian wave-packet state, $| \Psi_{in} \rangle|$ and the eigenstates, $|\tilde{\Psi}_n \rangle$ of the target Hamiltonians. We analyze the overlap distribution $|\langle \Psi_{in} | \tilde{\Psi}_n \rangle|^2$ superimposed on the IPR spectrum, as shown in Fig.~\ref{fig:8}(c)–(d). 
We show the spectrum of the logarithm of the IPR by orange color data points, while the overlap amplitudes are represented by symbol sizes proportional to their magnitudes, with colors corresponding to different wave-packet parameters as indicated in panels (a)–(b).

As shown in Fig.~\ref{fig:8}(c), for \(\tilde{H}_{\mathrm{axial}}\),  the overlap is broadly distributed over a large number of eigenstates with no discernible structure. This diffuse distribution implies contributions from a (quasi) continuum of energies, resulting in a wide range of incommensurate frequencies and rapid dephasing, thereby suppressing any revival of fidelity. As the wave-packet center is shifted towards lower $\langle \log(\mathrm{IPR}) \rangle$, an increasing number of eigenstates display a finite overlap amplitude. This further reinforces the suppression of any revival of fidelity for \(\tilde{H}_{\mathrm{axial}}\).
In contrast, for the largest block of \(\tilde{H}_{\mathrm{diag}}\), the overlap amplitude is sharply concentrated on a discrete set of eigenstates in the high-IPR towers (see Fig.~\ref{fig:8}(d)). When the wave-packet is centered at $\langle \log(\mathrm{IPR}) \rangle =-2.0$, it selectively overlaps with a small number of eigenstates that are nearly equally spaced in energy, leading to a limited set of coherent frequencies and hence the strong, nearly periodic revivals. As $\langle \log(\mathrm{IPR}) \rangle$ decreases, the fixed-width wave packet progressively overlaps with increasing number of eigenstates from a broader portion of these towers. This leads to an increase of the number of contributing frequencies and a gradual reduction of the revival amplitude. Further shifting the wave-packet center towards regions dominated by ergodic eigenstates, the overlap becomes increasingly defused, and the revivals are eventually washed out.

This analysis establishes a direct correspondence between spectral structure and wave-packet dynamics $-$ the presence of a regular, tower-like eigenstate structure enabling coherent revivals, whereas a diffuse, continuum-like spectrum leading to rapid dephasing. In particular, the selective overlap of the wave-packet with high-IPR tower states in the largest block of \(\tilde{H}_{\mathrm{diag}}\) identifies them as the primary carriers of scar-like dynamics, consistent with our previous study~\cite{Swain_EHC_QMC_PRB_2025}.

\begin{figure*}[t]
\centerline{
\includegraphics[width=0.9\linewidth]{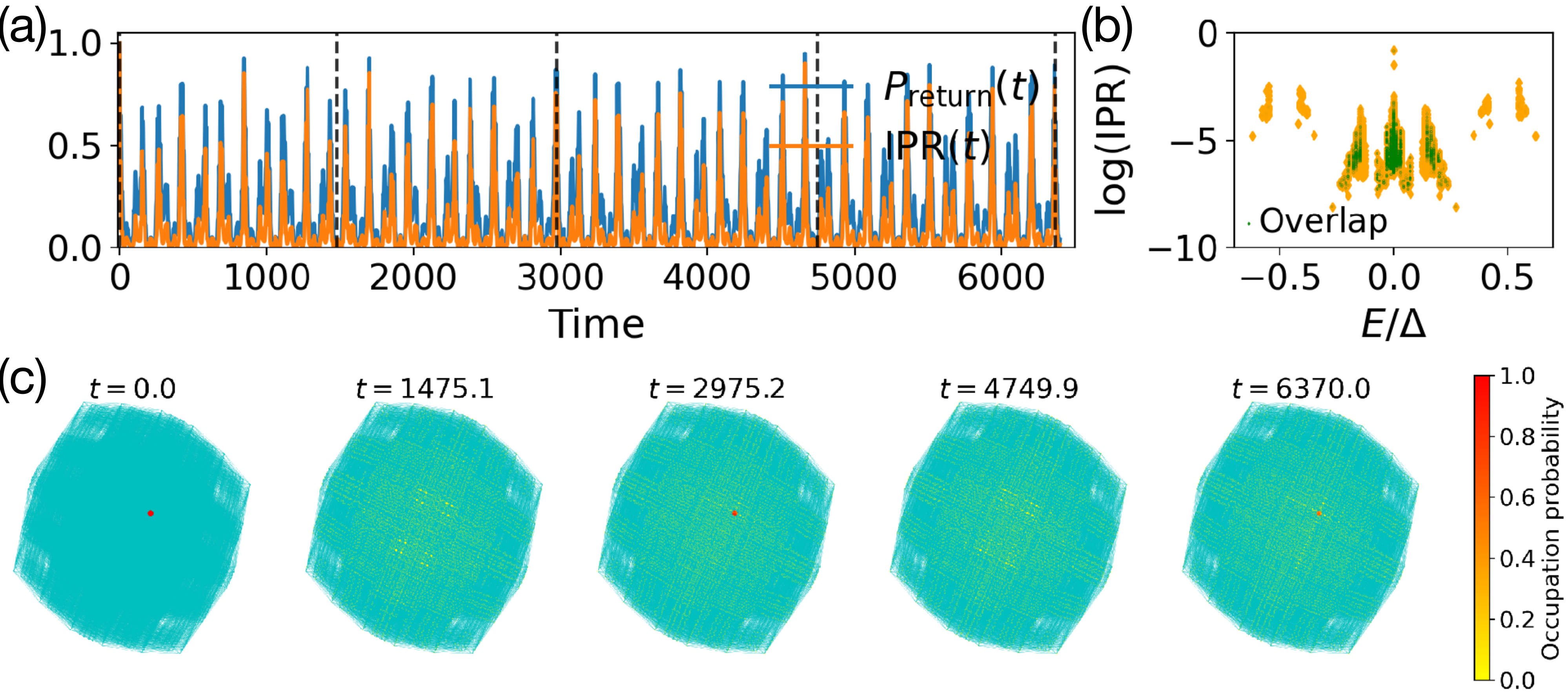}
}
\centerline{
\includegraphics[width=0.9\linewidth]{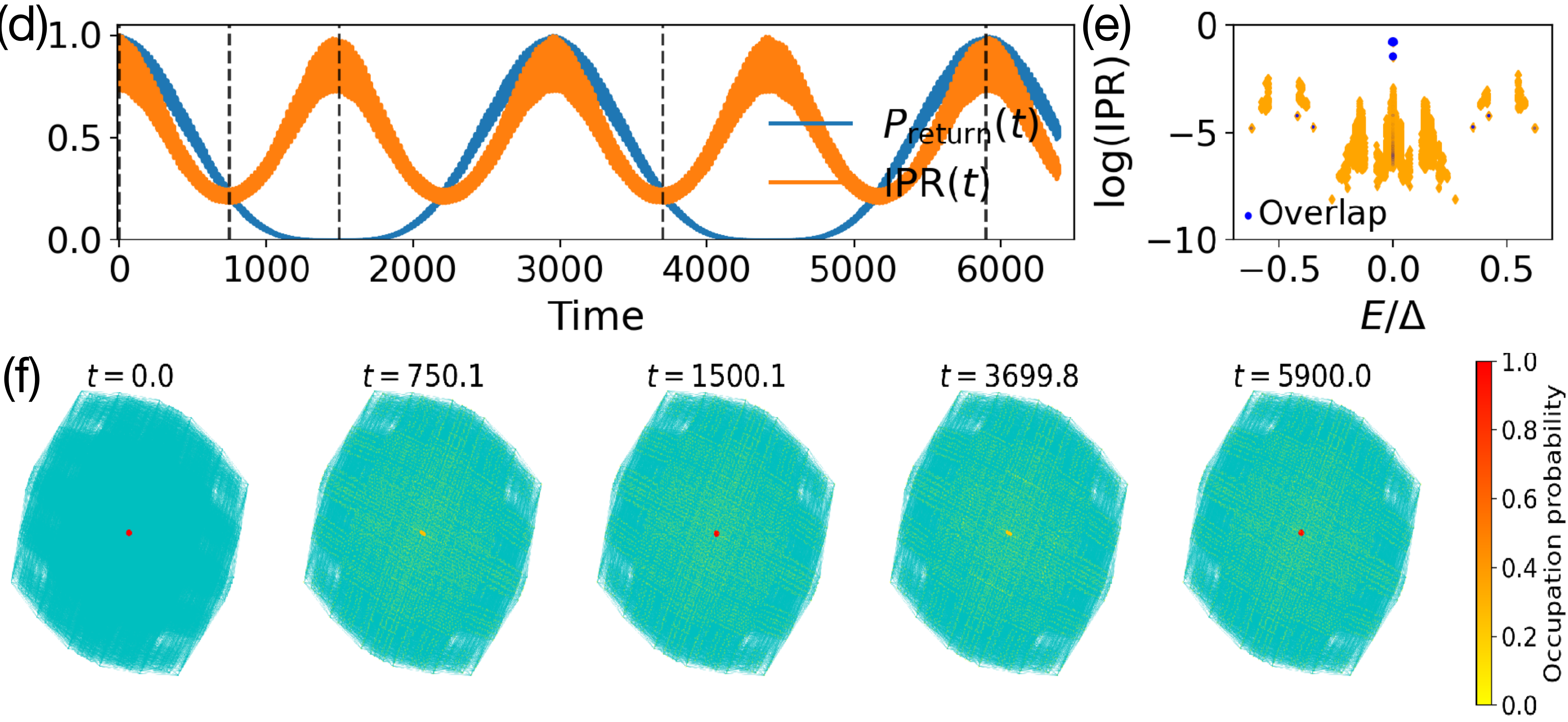}
}
\caption{\textbf{Fock-space dynamics for the largest block of \(\tilde{H}_{\mathrm{diag}}\) with \(\Delta=4\).}  Unitary time evolution of representative Fock-space basis states shows the role of Hilbert-space fragmentation and resonance structure in shaping revivals.  For the top panels (a)–(c), the initial state is characterized by $\mathrm{IPR}_{\alpha}\approx 0.002$ and node connectivity =16. 
(a)~The return probability and time-dependent IPR exhibit rapid, irregular oscillations. 
(b) The overlap distribution $|\langle \Psi_{\alpha} | \tilde{\Psi}_n \rangle|^2$, superimposed on the IPR spectrum, is spread over multiple eigenstates within the IPR towers, indicating participation of several frequencies. The logarithm of the IPR is shown in orange color data, while the overlap values are represented by green symbols with sizes proportional to the overlap magnitude.
(c) Fock-space snapshots show fast redistribution of probability confined within the subgraph, leading to high-frequency scar-like dynamics. 
The bottom panels (d)–(f) are for an initial state characterized by $\mathrm{IPR}_{\alpha}\approx 0.35$ and node connectivity=32. 
(d)~The return probability and IPR display a prominent coherent oscillations with a well-defined long revival period. 
(e)~The overlap superimposed on the IPR spectrum, is sharply concentrated on a few eigenstates with near-degenerate and regularly spaced energies, giving rise to discrete dynamical frequencies. The logarithm of the IPR is shown in orange color data, while the overlap values are represented by blue symbols with sizes proportional to the overlap magnitude.
(f)~Fock-space snapshots reveal a structured evolution of the basis state.}
\label{fig:10}
\end{figure*} 

\subsection{Fock space dynamics from localized basis state}

In generic many-body quantum systems, the structure of the Fock-space network has a profound influence on the nature of eigenstates and their dynamical evolution~\cite{Altshuler1997,Gornyi2005,Basko2006,multifractal_scaling_Laflorencie,KT_MBL_Gabriel2020,PhysRevB.108.075124,Scoquart2024FockSpace,Roy_Logan_FS_dynamics_JPCM2025}. 
Delocalized (ergodic) eigenstates typically explore a large fraction of the Fock-space network, with amplitudes spread across exponentially many basis states~\cite{Roy_Logan_FS_dynamics_JPCM2025}. Their time evolution rapidly equilibrates and the Fock-space probability distribution spreads diffusively across the network's high-connectivity graph~\cite{DeLuca2014}. 
By contrast, strongly localized eigenstates exhibiting high inverse participation ratios, occupy only a sparse fraction of Fock-space as seen in many-body localized systems~\cite{Altshuler1997,Gornyi2005,Basko2006,multifractal_scaling_Laflorencie,KT_MBL_Gabriel2020}. These localized states are robust under perturbations and evolve within a small subspace, leading to slow dynamics~\cite{Roy_Logan_FS_dynamics_JPCM2025}. 
Quantum many-body scar states represent a new nonthermal category, whose Fock-space profiles are neither extended nor exponentially localized~\cite{scar_states_Turner2018,scar_expt_driven_rydberg_2021,scar_states_Serbyn2021,Yao2023FockSpace2D}. Instead, they live on low-dimensional manifolds embedded within the full Fock-space graph, leading to partial/full Hilbert-space fragmentation~\cite{PhysRevB.108.075124,scar_rev_Moudgalya_2022}. This atypical structure yields nonergodic yet delocalized dynamics, including coherent oscillations and revivals, despite the surrounding thermal spectrum~\cite{rydberg_atom_2017,scar_states_Turner2018,Yao2023FockSpace2D}.

Fig.~\ref{fig:9} illustrates the unitary time evolution under \(\tilde{H}_{\mathrm{axial}}\) of two representative Fock-space basis states $-$ one of delocalized character ($IPR_{\alpha} \approx 0.005$ and node connectivity 20), described by panels (a)–(c), and another corresponding to localized character ($IPR_{\alpha} \approx 0.5$ and node connectivity 12), shown in panels (d)–(f). In both cases, the dynamics is initialized with the occupation probability fully concentrated on a single node of the Fock-space network. 
For the delocalized basis state, the return probability and the time-dependent IPR decay rapidly from their initial values, indicating efficient spreading over the Fock-space network [panel (a)]. The corresponding overlap distribution, $|\langle \Psi_{\alpha} | \tilde{\Psi}_n \rangle|^2$, shown in panel (b), is broadly distributed across a large number of eigenstates, consistent with strong hybridization. The Fock-space snapshots in panel (c) further reveal a rapid spread of occupation probability across the highly connected network, characteristic of ergodic dynamics.
In contrast, the localized basis state exhibits markedly different behavior. As shown in panel (d), both the return probability and the IPR remain close to their initial values even at long times, indicating persistent memory of the initial state. The overlap distribution is sharply concentrated on a small subset of eigenstates, reflecting limited hybridization. Consistently, the Fock-space snapshots in panel (f) show that, despite the high connectivity of the network, the occupation probability spreads only weakly and remains strongly localized near the initial node. This behavior signals non-ergodic dynamics arising from the structure of the underlying eigenstates. These results demonstrate that \(\tilde{H}_{\mathrm{axial}}\) supports coexisting ergodic and non-ergodic dynamical regimes, strongly depending  on the basis state.

In stark contrast to the axial model, the Fock-space dynamics under \(\tilde{H}_{\mathrm{diag}}\) exhibits clear signatures of scar-induced non-ergodic behavior. Fig.~\ref{fig:10} shows the time evolution of two representative Fock-space basis states from the largest block of \(\tilde{H}_{\mathrm{diag}}\).  
The first one is characterized by $IPR_{\alpha}\approx 0.002$, and node connectivity $=16$, and the other one is characterized by $IPR_{\alpha}\approx 0.35$ and node connectivity $=32$.
Because \(\tilde{H}_{\mathrm{diag}}\) decomposes the Hilbert space into disconnected sectors, the dynamics remains strictly confined to the subgraph containing the initial state. 
For the initial state characterized by $IPR_{\alpha}\approx 0.002$, and node connectivity =16 [Figs.~\ref{fig:10}(a)–(c)], the return probability exhibits rapid, irregular oscillations. The overlap distribution $|\langle \Psi_{\alpha} | \tilde{\Psi}_n \rangle|^2$, shown in panel (b), is distributed over multiple eigenstates within the IPR towers, 
indicating participation of several closely spaced energy levels. This produces fast dephasing followed by partial revivals. Correspondingly, the Fock-space snapshots [panel (c)] reveal rapid redistribution of probability within the subgraph, consistent with high-frequency revival dynamics characteristic of scar subspaces.

A qualitatively different behavior emerges for the initial state characterized by $IPR_{\alpha}\approx 0.35$ and node connectivity =32, as seen from
Figs.~\ref{fig:10}(d)–(f). In this case, the return probability and IPR display slower oscillations with a well-defined long revival period. The overlap distribution [panel (e)] is sharply concentrated on a small number of eigenstates, with dominant contributions arising from a pair of nearly degenerate eigenstates, supplemented by a few additional eigenstates with regular energy spacing. This structure gives rise to the long-period oscillatory dynamics. The corresponding Fock-space evolution [panel (f)] shows 
structured redistribution of probability within the subgraph without significant leakage.

Our results highlight the remarkable ability of the EHC framework to systematically design Hamiltonians with tailored Fock-space dynamics.  
With careful engineering of the interactions, \( \tilde{H}_{\mathrm{axial}} \) produces either fully ergodic or non-ergodic dynamics, whereas \( \tilde{H}_{\mathrm{diag}} \) produces periodic revivals characteristic of scar states. This capability is unique in the sense that it does not rely on fine-tuned constraints, disorder, or specific geometrical frustration. Consequently, EHC provides a powerful and flexible tool for realizing exotic quantum dynamics in two-dimensional spin systems.

\section{Conclusion}
\label{sec_concl}

In this work we establish the eigenstate-to-Hamiltonian construction as a systematic route to engineering tunable non-ergodic dynamics in two-dimensional quantum systems. By constructing two square-lattice Hamiltonians, \(\tilde{H}_{\mathrm{axial}}\) and \(\tilde{H}_{\mathrm{diag}}\) from the same input state but allowing different interaction geometries, we demonstrate how EHC enables engineering distinct Hilbert-space structures and many-body dynamics. 
In \(\tilde{H}_{\mathrm{axial}}\), the Hilbert space remains connected, yet increasing anisotropy suppresses resonant processes, driving a crossover from ergodic dynamics to a strongly non-ergodic regime, as consistently reflected in the IPR and entanglement entropy diagnostics, and the rapid loss of fidelity in wave-packet dynamics. In contrast, \(\tilde{H}_{\mathrm{diag}}\) realizes a qualitatively distinct mechanism $-$ Hilbert-space fragmentation combined with structured resonance patterns stabilize a hierarchy of low-entangled, high-IPR eigenstates forming tower-like spectral features. These states dominate dynamics, producing robust and tunable fidelity revivals, with their origin traced to selective overlap and near-regular energy spacings. A complementary analysis of the Fock-space dynamics uncovers fundamentally distinct dynamical regimes $-$ ergodic and non-ergodic behavior in the axial model, contrasted with confined and recurrent dynamics in the largest block of the diagonal model.

More broadly, our approach provides a blueprint for realizing non-ergodic dynamics in two-dimensional, disorder-free quantum systems~\cite{scars_2DPXP_2020_Lin,Yao2023FockSpace2D,Yuan2023HigherDimScars,Miao2025Exact2DScars}. 
Beyond their fundamental interest, quantum many-body scar states have recently emerged as valuable resources for quantum information processing. Their resistance to thermalization and coherent revival have enabled the generation of highly entangled states such as the Greenberger–Horne–Zeilinger (GHZ) states~\cite{Omran2019SchrodingerCat}, as well as protocols for enhancing entanglement far from equilibrium~\cite{Dong2023DisorderEntanglement}. These properties have further motivated proposals for robust quantum sensing and entanglement-enhanced metrology in scar systems~\cite{Dooley2021RobustSensing,Desaules2022Multipartite,Dooley2023EntanglementMetrology,Papic2022EntanglementSpinChains}. 
In this context, the ability to systematically engineer and tune scar-hosting Hamiltonians, as demonstrated here, opens new avenues for designing platforms with potential applications in quantum technologies. Importantly, our framework is compatible with programmable quantum simulators — such as Rydberg atom arrays and superconducting architectures — where interactions and constraints can be precisely tailored~\cite{scar_expt_driven_rydberg_2021,Zhang_scar_expt_qcomp_2023,Gustafson_scar_qcomp_2023}. This provides a promising pathway for experimentally realizing engineered non-ergodic phases and scar-like dynamics beyond one dimension.

\begin{figure*}[t]
\setcounter{figure}{0}
\renewcommand{\thefigure}{A\arabic{figure}}
\centerline{
\includegraphics[width=0.9\linewidth]{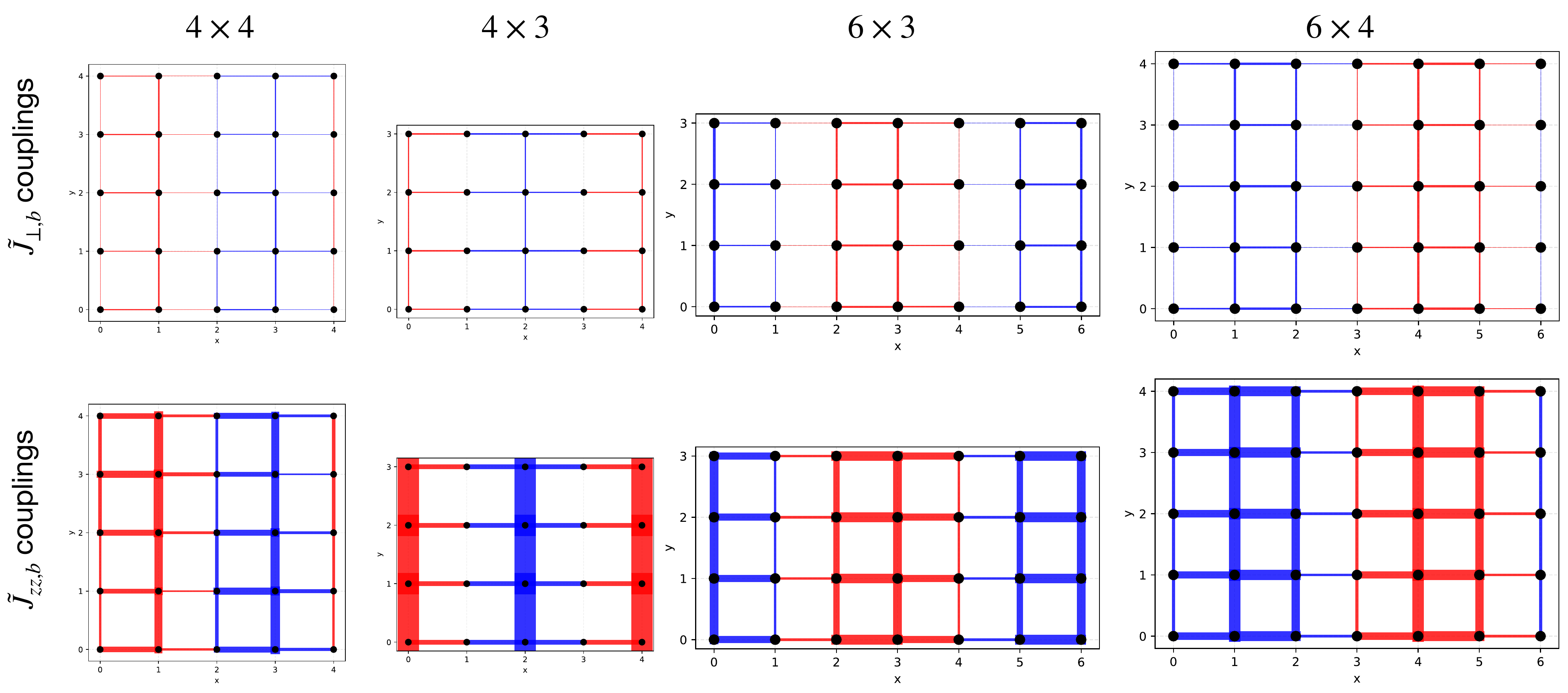}
}
\centerline{
\includegraphics[width=0.9\linewidth]{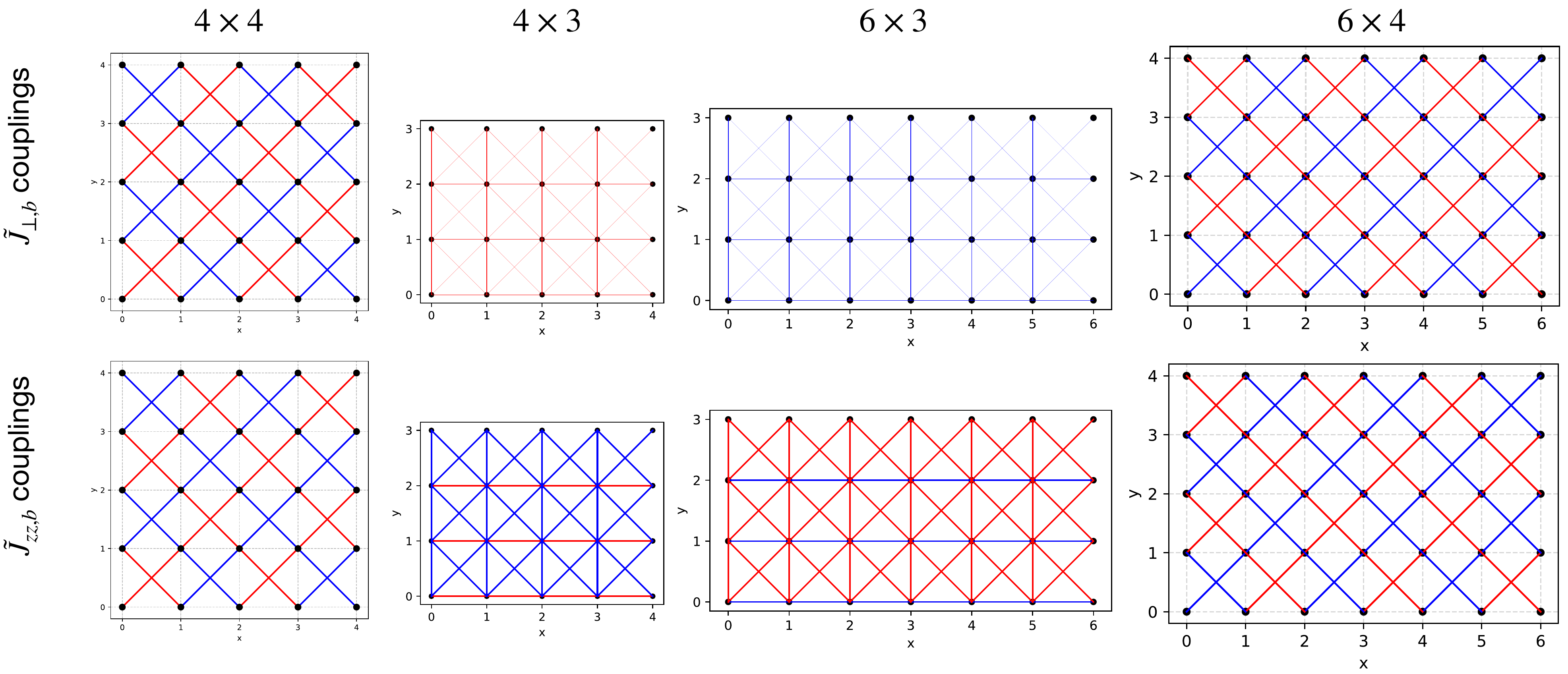}
}
\caption{
\textbf{Coupling patterns of \(\tilde{H}_{\mathrm{axial}}\) and  \(\tilde{H}_{\mathrm{diag}}\) across different system sizes.}
Spatial distribution of the spin-flip and spin-preserving couplings obtained from the EHC procedure on \(4 \times 4\), \(4 \times 3\), \(6 \times 3\), and \(6 \times 4\) lattices. Bond thickness denotes coupling magnitude, and color indicates antiferromagnetic (red) or ferromagnetic (blue) interactions.
The top two rows correspond to the couplings of \(\tilde{H}_{\mathrm{axial}}\), which exhibits structures that closely mirror those of the \(4 \times 4\) lattice across all system sizes, indicating strong geometric robustness.
The bottom two rows correspond to the couplings of \(\tilde{H}_{\mathrm{diag}}\), where the coupling patterns on \(4 \times 3\) and \(6 \times 3\) differ from the \(4 \times 4\) case, while the \(6 \times 4\) lattice faithfully reproduces the coupling structures observed on the \(4 \times 4\) case. 
} 
\label{fig:11}
\end{figure*} 

\begin{acknowledgements} 

We thank Pinaki Sengupta, Ho-Kin Tang, Darryl Chuan Wei Foo, Fakher F. Assaad, Kai Dieckmann, and Mohitha Adira for helpful discussions.  This work is supported by the Singapore Ministry of Education AcRF Tier 2 grant MOE-T2EP50222-0005.  We also acknowledge the use of computational resources at the Singapore National Super Computing Centre (NSCC) ASPIRE-2A cluster, at Calcul en Midi-Pyrénées (CALMIP) and NUS IT's Research Computing group under grant number NUSREC-HPC-00001 for our simulations. SA thanks support from the Singapore National Research Foundation Investigator Award (NRF-NRFI06-2020-0003) and a start-up grant from Washington University in St. Louis; GL acknowledges Singapore Ministry of Education (MOE-T2EP50223-0009), and the EUR Grant NanoX No. ANR-17-EURE0009 in the framework of the “Programme des Investissements d'Avenir”, by France 2030 under the French National Research Agency Award QUTISYM No. ANR- 23-PETQ-0002 and by the ANR Research Grants ManyBodyNet No. ANR-24-CE30-5851.

\end{acknowledgements}

\appendix

\begin{figure*}[t]
\setcounter{figure}{1}
\renewcommand{\thefigure}{A\arabic{figure}}
\centerline{
\includegraphics[width=0.9\linewidth]{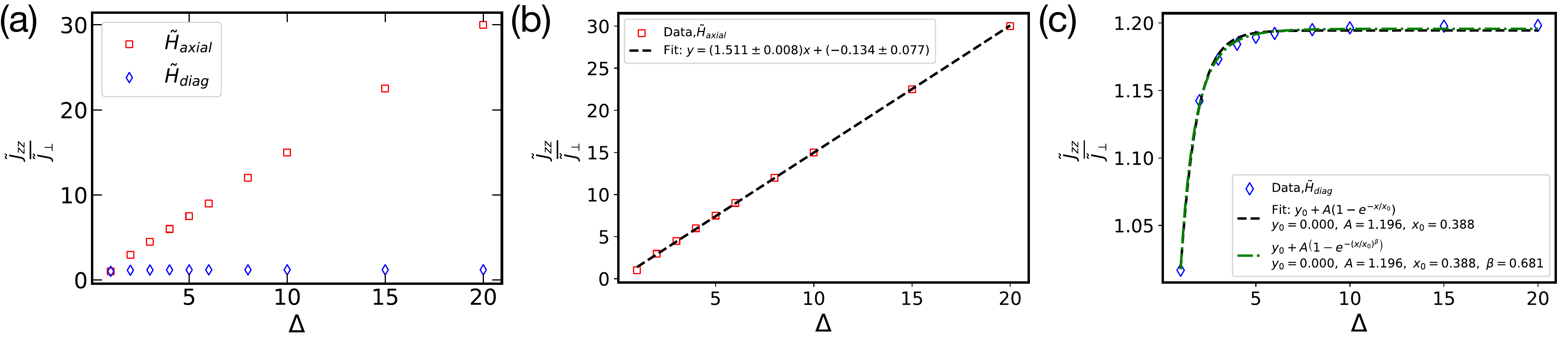}
}
\caption{
\textbf{Dependence of the effective coupling ratio $J_{zz}/J_{\perp}$ on $\Delta$ for the EHC models.}
(a) Evolution of the ratio with $\Delta$ for the axial and diagonal models. The axial model exhibits a smooth monotonic increase, whereas the diagonal model shows a rapid initial variation followed by saturation.
(b) Linear fit to the axial-model data, demonstrating an approximately linear dependence on $\Delta$ over the studied range.
(c) Exponential and stretched-exponential fits to the diagonal-model data, capturing its nonlinear approach to saturation. The corresponding fitting parameters are indicated in the panels.
} 
\label{fig:12}
\end{figure*} 

\section{Robustness of Engineered Hamiltonians Across System Sizes}

In the main text, we focus on the EHC built target Hamiltonians $-$ \(H_{\mathrm{axial}}\) and \( H_{\mathrm{diag}}\) on square lattice of size \(4 \times 4\). Our study involves a detailed characterization of the non-ergodic properties of these target Hamiltonians via spectral and dynamical properties using exact diagonalization method. To demonstrate that our conclusions are not specific to this choice of system size, in this appendix we present both the axial and diagonal target Hamiltonians on \(4 \times 3\), \(6 \times 3\) and \(6 \times 4\) lattices and compare their coupling patterns with those obtained for the \(4 \times 4\) system.

Across all lattice sizes studied, the EHC procedure produces \(H_{\mathrm{axial}}\) with nearest-neighbor coupling patterns that are structurally equivalent up to boundary effects, and closely mirror those obtained for size \(4 \times 4\), despite the increased Hilbert-space dimension. The resulting couplings are shown in fig.~\ref{fig:11}. Thus, for the axial model, we find that the EHC is remarkably insensitive to lattice geometry.

In contrast, the diagonal model built with the EHC procedure, displays a pronounced sensitivity to lattice geometry. When it is applied to \(4 \times 3\) and \(6 \times 3\) lattices, the couplings of resulting target Hamiltonian differ qualitatively from those of the \(H_{\mathrm{diag}}\) on lattice \(4 \times 4\). In these geometries, the interactions reorganize in a manner that reflects the reduced lattice symmetry, leading to modified spatial patterns of the couplings. However, when the EHC is carried out on a \(6 \times 4\) lattice, whose geometry closely resembles that of the \(4 \times 4\) lattice, we recover the coupling structures of \(H_{\mathrm{diag}}\) that are closely analogous to those observed in the main text (see figure). The geometric dependence of the coupling patterns thus reflects the intrinsic sensitivity of diagonal interactions to lattice symmetry, rather than a breakdown of the EHC approach.

\section{Dependence of effective coupling ratios on the XXZ anisotropy}
\label{app_B}

Our EHC models are characterized by spin-flipping ($J_{\perp}$) and spin-preserving ($J_{zz}$) interactions defined on nearest-neighbor axial or next-nearest-neighbor diagonal bonds of the square lattice. As we do not fix the ratio $J_{zz}/J_{\perp}$ a priori in EHC, it is important to determine how this ratio evolves with the XXZ anisotropy parameter, $\Delta$ across the regimes studied in the main text. We show our results in fig.~\ref{fig:12}.

For the axial model, the ratio $J_{zz}/J_{\perp}$ increases monotonically with $\Delta$ (see panel (b)). This smooth evolution is consistent with a gradual crossover from conventional thermalization to increasingly constrained dynamics identified in our spectral and dynamical diagnostics. In contrast, the diagonal model exhibits a strongly nonlinear dependence of $J_{zz}/J_{\perp}$  on $\Delta$. The ratio changes rapidly at small anisotropy and subsequently approaches a limiting value at larger $\Delta$, consistent with an exponential or stretched-exponential saturation behavior (see panel (c)).

\section{Studying 2D scars with QMC}
In Ref.~\cite{Swain_EHC_QMC_PRB_2025} we provided a framework for engineering many-body quantum Hamiltonians with nonergodic properties using the Quantum Monte Carlo (QMC) method. This approach combines the EHC scheme with QMC sampling, known as the EHC-QMC method, enables the design of Hamiltonians that host quantum many-body scar states. Specifically, we demonstrated that in one-dimensional spin-half XXZ chains, one can systematically construct target Hamiltonians hosting scar states $-$ within a computationally tractable QMC setup.

Extending this EHC-QMC approach to two-dimensional systems introduces additional complexity. For the axial model $\tilde{H}_{\mathrm{axial}}$ that does not host scar states this QMC-based implementation remains feasible because the Hamiltonian involves only nearest-neighbor (NN) antiferromagnetic couplings along the principal lattice axes, thus avoiding the sign problem in QMC. 
In contrast, the model $\tilde{H}_{\mathrm{diag}}$, which hosts quantum many-body scar states, necessarily incorporates next-nearest-neighbor antiferromagnetic interactions along the lattice diagonals. The coexistence of NN and NNN antiferromagnetic couplings introduces geometric frustration, resulting in negative path-integral weights and consequently a severe sign problem that renders QMC simulations exponentially difficult. This limitation prevents a straightforward extension of the EHC-QMC method to build such scar model in 2D. 

One possible remedy is to replace the NNN antiferromagnetic interactions with weak ferromagnetic ones while keeping the NN couplings antiferromagnetic. This modification can mitigate the sign problem and may allow an approximate exploration of the target Hamiltonians in 2D, thereby extending the applicability of the EHC-QMC framework beyond its current 1D domain to build models hosting scar states.

\newpage

\bibliography{references}

\end{document}